\documentclass[preprintnumbers,prd,floatfix,superscriptaddress,nofootinbib,notitlepage]{revtex4-1}
\pdfoutput=1
\usepackage{graphicx,epsfig}
\usepackage{amsmath}
\usepackage {amssymb}
\usepackage {longtable}
\usepackage{multirow}
\usepackage{dcolumn}
\usepackage{bm}
\usepackage{amsfonts}
\usepackage{subfigure}
\usepackage{color}
\usepackage{relsize}

\begin{document}

\title{Penrose process in Reissner-Nordstr\"om-AdS black hole
spacetimes:\\ Black hole energy factories and black hole bombs}

\author{Duarte Feiteira}
\email{duartefeiteira@tecnico.ulisboa.pt}
\affiliation{Centro de Astrof\'{\i}sica e Gravita\c c\~ao  - CENTRA,
Departamento de F\'{\i}sica, Instituto Superior T\'ecnico - IST,
Universidade de Lisboa - UL, Av. Rovisco Pais 1, 1049-001
Lisboa, Portugal}

\author{Jos\'e P. S. Lemos}
\email{joselemos@ist.utl.pt}
\affiliation{Centro de Astrof\'{\i}sica e Gravita\c c\~ao  - CENTRA,
Departamento de F\'{\i}sica, Instituto Superior T\'ecnico - IST,
Universidade de Lisboa - UL, Av. Rovisco Pais 1, 1049-001
Lisboa, Portugal}

\author{Oleg B. Zaslavskii}
\email{zaslav@ukr.net}
\affiliation{Department of Physics and Technology,
Kharkov V.~N.~Karazin National
University, 4 Svoboda Square, 61022 Kharkov, Ukraine}

\begin{abstract} 
The Penrose process for the decay of electrically charged particles in
a Reissner-Nordstr\"om-anti-de Sitter black hole spacetime is studied.
To extract large quantities of energy one needs to mount a recursive
Penrose process where particles are confined and can bounce back to
suffer ever again a decaying process in the black hole electric
ergoregion.  In an asymptotically anti-de Sitter (AdS) spacetime, two
situations of confinement are possible.  One situation uses a
reflecting mirror at some radius, which obliges the energetic outgoing
particles to return to the decaying point. The other situation uses
the natural AdS property that sends back at some intrinsic returning
radius those outgoing energetic particles.  In addition, besides the
conservation laws the decaying process must obey, one has to set
conditions at the decaying point for the particles debris. These
conditions restrain the possible scenarios, but there are still a
great number of available scenarios for the decays.  Within these, we
choose two scenarios, scenario 1 and scenario 2, that pertain to the
masses and electric charges of the final particles.  Thus, in the
mirror situation we find that scenario 1 leads to a black hole energy
factory, and scenario 2 ends in a black hole bomb. In the no mirror
situation, i.e., pure Reissner-Nordstr\"om-AdS, scenario 1 leads again
to a black hole energy factory, but scenario 2 yields no bomb. This
happens because the volume in which the particles are confined
increases to infinity along the chain of decays, leading to a zero
value of the extracted energy per unit volume and the bomb is demined.
The whole treatment performed here involves no backreaction on the
black hole mass and black hole electric charge, nevertheless we
speculate that the end state of the recursive process is a
Reissner-Nordstr\"om-AdS black hole with very short hair, i.e., with
one particle at rest at some definite radius.
\end{abstract}
\maketitle

\section{Introduction}
\label{sec:intro}

The Penrose process was the first process discovered to extract energy
from a black hole, specifically, from a
rotating Kerr black hole \cite{penrose_book,penrosefloyd}. It uses
the decay of a particle into other particles at the black hole
ergoregion, a region with negative energy states relatively to
infinity. In the decay of the initial particle it can happen that one
of the new particles falls into the black hole and the other particle
escapes to infinity. The particle falling into the black hole, since
it crosses the ergoregion, has negative energy, and due to
energy conservation in the decay, the escaping particle must have a
larger energy than the original particle. The source of energy
extraction mechanism is the angular momentum of the black hole. The
Penrose process yielded dynamic ways to probe a black hole and indeed
triggered the train of ideas that led to the four laws of black hole
mechanics.  Since the net energy extraction efficiency of the Penrose
process is bounded by an upper limit of the order of twenty percent
\cite{bardeen,wald}, one can think of other mechanisms, different from
decays, to improve the performance at the ergoregion.  One such
mechanism is through particle collisions, although it was found that
there are not robust improvements over the original process
\cite{psk}. However, it was found that when the collision of two
particles is near the horizon of a maximally rotating black hole,
i.e., an extremal black hole, one has that the center of mass energy
is divergingly high if the angular momentum of the particles is
critically tuned, a result called the BSW effect \cite{bsw}.  Thus,
such extremal black holes can create highly energetic particles or
even highly massive ones.  It was further shown that, if instead of an
exactly critical particle, a near-critical particle is used, and the
parameters are specially adjusted at the point of collision, the
effect subsists \cite{grib}.  Some issues persist, for instance, for
nonextremal rotating black holes, a centrifugal barrier impedes the
critical particle from reaching the horizon \cite{gao}, and indeed one
can set upper bounds on the energy of the emitted particle and on the
extraction efficiency for the BSW effect and generalization of it
\cite{zalavskii2,harada}.  For reviews, see
\cite{haradakimura,schnittman}.
For collisions and extraction of energy in generic
rotating sources
with the possibility of having
energy extraction without bounds
see \cite{zaslavskii3},
and for
a
rather general treatment of the processes
in the equatorial plane of a
rotating black hole see \cite{new,newnew,zalavskii7}.
For static electrically charged black holes there also exists an
ergoregion where the energy of electric charged particles can be
negative and so energy can be extracted from the black hole in a
similar way to the rotating case \cite{denardo}. In this case, the
energy source is the Coulomb energy of the black hole.  Moreover,
energy extraction from a black hole with an electric field can be more
efficient than that from a neutral rotating black hole as the
efficiency of the process in the electric case is not bounded
\cite{wdd}.  In addition, the counterpart of the BSW collision effect
exists for extremal electrically charged black holes
\cite{zalavskii1}, and
a collision of particles can yield arbitrarily high values for
the escaping particle mass or energy at infinity if one of the
particles has a fine-tuned value of the electric charge, a phenomenon
that does not happen for a rotating black hole case.  So, there is no
upper bound on the energy extraction from these collisions, and the
process is referred to as a super Penrose process
\cite{zalavskii5,nemoto,zalavskii6}.

Another way to extract energy from a black hole, namely,
through waves, is a process called superradiance. Instead of particles,
as in the Penrose process, in superradiance one sends a wave field
into the ergoregion, to get back, under certain conditions, an
amplified wave.  Superradiation for a rotation object was considered
in \cite{zeldovich} where it was examined what happens when scalar
waves hit a rotating cylinder.  Considering a wave incident upon such
a rotating object, one finds that if a measure of the frequency of the
incident wave is smaller than the angular velocity of the body, then
the scattered wave is amplified. A Kerr black hole, being the
paradigm of rotating objects in general relativity, certainly
displays superradiant phenomena \cite{starob,teukpress1}.
One can also have superradiance from
electrically charged waves incident upon
an electrically charged black hole, in this case the condition for
amplification of the scattered wave is
that the frequency of the incident
wave be smaller than the product of the electric charge of the wave
times the electric potential \cite{dr}.

There is now the prospect that the amplified scattered wave may be
fed back to the black hole in a recursive process, so that one can
extract as much rotational energy as possible from a Kerr rotating
black hole.  For instance, if one surrounds the black hole by a
reflecting mirror, the amplified wave will bounce in and out between
the mirror and the black hole ergoregion with the result that it is
itself amplified each time.  The total energy extracted from the black
hole grows exponentially until in some way or another the radiation
pressure wipes out
and destroys the mirror in a run away explosive event
or, the black hole exhausts, in some way or another, its source of
energy.  This exponential growth of the energy extracted is the black
hole bomb \cite{teukpress2},
and its equivalent for any rotating object has also
been displayed \cite{vilenkin}.
Detailed models for the black hole bomb
have
been proposed, notably one in which a scalar field wave has been used
\cite{cdly}, where it was found that for the system to work as a bomb
there is a minimum distance at which the mirror must be located.
Superradiance
and bombs in analog gravity black hole models have also
been shown to set in, see 
\cite{lemosanalogue} for a review.
It
is by now clear that a black hole bomb can also be installed in
an electrically charged black hole, e.g., a Reissner-Nordstr\"om black
hole.  By adding a reflecting mirror suitably far away from the black
hole horizon, electrically charged waves can be amplified essentially
without limit to extract energy from the black hole
\cite{dhr,dolanpw,menzay}.
Conversely,
it is also possible to argue that
one can obtain a black hole bomb through the Penrose
process, instead of through superradiance.
For that, one must confine the outgoing energetic particles by
placing a reflecting mirror at some point of the
space. This confinement
has been done in an ingenious manner for electrically
charged particles in a 
Reissner-Nordstr\"om black hole
spacetime
\cite{penrosekok}.  Then, one has to set
specific conditions that the
escaping particles, that return to the decaying point, must satisfy such
that the recursive Penrose process with the sequence of particle
decays can make a black hole bomb,
as has been done in a rotating
black hole
background and in an electric charged black hole background
\cite{penrosezasl}.  

One way of disposing of the reflecting mirror, which is effective in
setting the instability and trigger the bomb, is by considering a
scalar field with mass scattering off a Kerr black hole, as the
field's mass acts effectively as a mirror
\cite{ddr,furuhashinambu}, or by considering
a massless
scalar field in a
Kerr black hole spacetime with small extra dimensions
attached to it,  with these
giving in fact
a mass to the  field and thus providing a natural mirror \cite{cl}.
Another way is to consider an
asymptotically anti-de Sitter (AdS) spacetime, since its negative
cosmological constant functions efficiently as a box, i.e., a reflective
mirror.  For instance, the results given in \cite{cdly} for a black
hole with a mirror enabled giving arguments to understand the reason
that small Kerr-AdS black holes are unstable and large Kerr-AdS black
holes are stable to superradiance.  Now, when an instability sets in,
one should look for the endpoint of the unstable system, which means
that backreaction effects must be considered.  In the electric
charged case, it has been known that linear instabilities that
appear for charged scalar fields in Reissner-Nordstr\"om-AdS black
holes evolve nonlinearly, through backreaction, to a hairy black
hole, and, interestignly enough, these are related via the AdS/CFT
correspondence to some form of superconducting transition
phenomena
\cite{hawkingr,gubser,hartnoll,uchikatayoshida,green,sanchis,bosch}.
Given that waves, in the form of superradiance, in
Reissner-Nordstr\"om-AdS black hole spacetimes, present instabilities,
bombs, and many other features, it is of interest, to understand
whether the Penrose process by itself in a recursive way yields
instabilities with the corresponding black hole bomb in those same
spacetimes. In the case the instability sets in it would be interesting
to ponder whether backreaction might lead to a black hole with hair.

Our aim is
to investigate the recursive Penrose process in a
Reissner-Nordstr\"om-AdS black hole spacetime, i.e.,
in an electrically
charged black hole spacetime with negative cosmological constant,
and analyze the possibility of creating a black hole bomb. We
deal with two situations, in the first there is a spherical reflective
mirror placed at some radius outside the event horizon, in the second
there is no mirror, i.e., we consider the Penrose process in a pure
Reissner-Nordstr\"om-AdS black hole spacetime.  In this second
situation, it is used the fact that a particle can remain confined in
a region between two turning points, which is important for a
recursive Penrose process.  As we will see, within the two situations,
one with mirror, the other with no mirror, there are cases that give
black hole energy factories, there are other cases that give black
hole bombs, and there are cases that yield no effective output.
In \cite{chandrasekhar} and \cite{olivaresetal}
the motion of
charged particles on Reissner-Nordström
and Reissner-Nordström-AdS black
hole spacetimes, respectively, has been given.

The work is organized as follows. In Sec.~\ref{sec:equations}, the
main equations describing charged particle motion in a
Reissner-Nordstr\"om-AdS black hole spacetime are presented, as well
as the two possible situations of interest, i.e., the mirror
situation, and the no mirror situation. In Sec.~\ref{sec:conditions},
the important assumptions related to the the decays of the particles
are stated, and the displaying of two possible decaying scenarios is
considered. In Sec.~\ref{sec:confined}, in the presence of a
reflective mirror, the two scenarios are analyzed which lead to black
hole energy factories and black hole bombs. In
Sec.~\ref{sec:no_mirror}, in a pure Reissner-Nordstr\"om-AdS black
hole spacetime, i.e., a spacetime with no mirror, the two scenarios
are analyzed which lead to black hole energy factories and no black
hole bombs, with the analysis focusing in energy extraction aspects
and particle motion between the turning points, and giving the mass,
charge, energy, and position of the outer turning point for even and
odd particles after infinite decays in the no mirror situation
succinctly in a table.  In Sec.~\ref{sec:concl}, we conclude,
contemplating what can happen if backreactions effects taken into
consideration.  Several appendices are needed to complement the work.
In Appendix \ref{hradii}, we determine the characteristic radii of the
Reissner-Nordstr\"om-AdS black hole spacetime in terms of the
spacetime mass, electric charge, and cosmological length.
In Appendix \ref{interesting}, a convenient compact formula for the
conserved quantities is given.  In Appendix \ref{sec:odd_dynamics},
the dynamics of particles moving inwards is analyzed.  In Appendix
\ref{sec:turning_position}, the position of the outer turning point in
the decaying scenarios 1 and 2 is found.  In Appendix
\ref{sec:volume}, the volume of the region between the event horizon
and some definite radius in the Reissner-Nordstr\"om-AdS spacetime is
calculated, and the volume of the region in Reissner-Nordstr\"om-AdS
spacetime with turning point radius in decaying scenario 2 when the
number of decays goes to infinity is found.  We use geometric units in
which the gravitational constant $G$, and the speed of light $c$ are
set to one, $G=1$ and $c=1$.

\section{Line element, equations of motion, turning points, and
situations of confinement for the Penrose process}
\label{sec:equations}

\subsection{Line element}

The line element for the spacetime interval $s$
in the Reissner-Nordstr\"om-AdS black hole spacetime
in $(t,r,\theta,\phi)$ spherical coordinates is given by
\begin{equation}
 ds^2 = - f\left(r\right) dt^2 +
\frac{dr^2}{f\left(r\right)} + r^2
\left( d\theta^2+\sin^2\theta\, d\phi^2\right),
\quad\quad\quad
f\left(r\right) = \frac{r^2}{l^2} + 1 -
\frac{2M}{r} +\frac{Q^2}{r^2}\,,
 \label{eq:f_RN_AdS}
\end{equation}
where
$M$ is the mass of the black hole
spacetime, $Q$ is its electric charge,
$l^2 = - \frac3{\Lambda}$ is
the length scale related with the negative
cosmological constant $\Lambda$,
and $f(r)$ is the metric potential. 
The coordinate ranges are $-\infty<t<\infty$, $r_+<r<\infty$,
$0\leq\theta\leq\pi$, and $0\leq\phi<2\pi$, where
the radius $r_+$ is
the black hole event horizon radius that is found from the equation
$f(r)=0$, so that 
\begin{equation}
r_+=r_+(M,Q,l)\,,
\label{eq:r+def}
\end{equation}
for some definite expression of
$r_+(M,Q,l)$.
There is another physical radius, $r_-$, for which
$f(r)=0$. In terms of these
radii,
$f(r)$ can be put in the form
$f(r)= \frac{1}{r^2}(r-r_+)(r-r_-)\left(\frac{r^2}{l^2} +
\frac{r_+ + r_-}{l^2}
r + 1+\frac{r_+^2 + r_-^2 + r_+ r_-}{l^2}\right)$,
see Appendix \ref{hradii}
for the 
expressions of $r_+$ and $r_-$
in terms of $M$, $Q$, and $l$.
We are interested
in the region $r>r_+$, so $r_-$ does not
play a role.
The electric potential $\varphi$ of the spacetime is 
\begin{equation}
\varphi=\frac{Q}{r}\,,
\label{eq:electricpot}
\end{equation}
i.e., a Coulomb electric potential.

\subsection{Equations of motion}

The equations of motion for a particle with mass $m$ and
specific charge $\tilde e$,
moving in the
Reissner-Nordstr\"om-AdS
spacetime, can be obtained from a Lagrangian formalism,
considering the Lagrangian $L$ with a Coulomb interaction term
given by
$2L = - f\left(r\right) \dot{t}^2 +
\frac{\dot{r}^2}{f\left(r\right)} + r^2 \dot{\phi}^2 -
\frac{2 {\tilde e} Q}{r} \dot{t}$,
where a dot means a derivative relative to the
particle's
proper time $\tau$, defined through $d\tau^2=-ds^2$.
From the Euler-Lagrange equations for $L$ one obtains
the equations of motion. 
The equation of motion for
the particle's coordinate $t$ is $\frac{d \;}{d
\tau}\left(\frac{\partial L}{\partial\dot t}\right)=0$, i.e.,
$\frac{d \;}{d \tau}\left(f{\dot t}+\frac{ {\tilde e} Q}{r}
\right)=0$, which has a first integral given by $f{\dot t}+\frac{
{\tilde e} Q}{r} =\frac{E}{m}$, where $E$ is the energy of the
particle, a conserved quantity. Thus, ${\dot t}=\frac{E-\frac{
e Q}{r}}{mf}$, where
the particle's total electric charge has been defined
as $e=m\tilde e$. Assuming radial motion for the particle, one
finds that the first integral
for the radial coordinate is 
$\dot{r}^2 = \frac{\left( E-\frac{ e  Q}{r}\right)^2}{m^2}-f$.
Defining the time component of the 4-momentum of the particle
as $p^t\equiv m \dot{t}$ and the
radial  component of the momentum of the particle
as $p^r\equiv m \dot{r}$ one then finds that the
equations of motion for the
particle can be put in the form
\begin{equation}
p^t\equiv m \dot{t} = \frac{X}{f},
\quad\quad\quad\quad
p^r \equiv m \dot{r} = \sigma P\,,
\label{eq:motion_t_penrose}
\end{equation}
where the quantities $X$ and $P$ are defined as
\begin{equation}
X =   E - \frac{ e Q}{r}, \quad\quad\quad\quad P =
  \sqrt{X^2 - m^2 f},
    \label{eq:motion_r_penrose}
\end{equation}
with $f=f(r)$ being given in Eq.~\eqref{eq:f_RN_AdS},
and $\sigma = \pm 1$ defining the direction of
the particle's motion,
$+1$ for outward radial  motion and $-1$
for inward motion.
The forward in time condition, $\dot{t} > 0$
implies that $X > 0$ outside the horizon.

\subsection{Turning points}

The turning points $r_{\rm t}$
of the particle's motion
are given by
the equation
\begin{equation}
\dot r=0\,,\quad\quad{\rm at}\quad r=r_{\rm t}\,,
    \label{eq:turning_points_generic}
\end{equation}
i.e., $P = 0$.
Using Eq.~\eqref{eq:motion_r_penrose}
one finds 
\begin{equation}
 - \frac{r_{\rm t}^4}{l^2} + \left(\frac{E^2}{m^2} - 1\right)
 r_{\rm t}^2 + 2 \left(1 - 
 \frac{EeQ}{m^2M}\right) M\,r_{\rm t} +
 \left(\frac{e^2}{m^2} - 1\right) Q^2 = 0,
    \label{eq:turning_points_penrose}
\end{equation}
In the presence of
a negative
cosmological
constant there are
two turning points outside the event horizon, the
inner turning point which  we call $r_{\rm i}$
and the outer turning point which
we keep its notation as $r_{\rm t}$. 
In the case of zero cosmological constant, i.e., pure
Reissner-Nordstr\"om black hole spacetimes, the equation for the
turning points gives only one solution, i.e., there is only one
turning point outside the horizon.

\subsection{Situations of confinement for recursive Penrose processes in
Reissner-Nordstr\"om-AdS black hole spacetimes: Confinement with
reflective mirror and with no reflective mirror}

Since in Reissner-Nordstr\"om black hole spacetime without
cosmological constant there is only one turning point, it was assumed
in \cite{penrosekok,penrosezasl} that the decay needed for Penrose
process happened at this single turning point $r_{\rm i}$.  But, as the
emerging new 
particles travel outward, in order to have multiple Penrose processes
one is obliged to put a
reflective spherical mirror at some outer radius $r_{\rm m}$ to
return back the even particles, such that there is confinement for
all processes \cite{penrosezasl}.

In the presence of a negative cosmological constant, in
Reissner-Nordstr\"om-AdS black hole spacetimes, there are two turning
points outside the event horizon, which indeed are solutions of
Eq.~\eqref{eq:turning_points_penrose}, one is 
$r_{\rm i}$ the other is $r_{\rm t}$.
As the emerging 
particles travel outward,
in order to have multiple Penrose processes, one can also
have a situation
where a reflective spherical mirror is put at some
radius $r_{\rm m}$ to reflect back the outgoing particles
to yield confinement.
However, there is an alternative to the reflective mirror
that also yields confinement. Indeed,
since in a
Reissner-Nordstr\"om-AdS black hole spacetime
there is the second outer turning point $r_{\rm t}$, this very
turning point can replace the role of the reflective mirror to allow
a situation
where multiple decays are still
possible and yield a consequent recursive Penrose process.
One subtlety that arises though in this situation,
is that the position of this outer
turning point depends on the ratio between the charge and the mass of
the particle as well as on the ratio between its energy and its mass,
and
in the case of interest
moves outwards. This fact introduces subtleties.

\section{Conditions for decay and the two different decaying scenarios}
\label{sec:conditions}

\subsection{Conditions}

\subsubsection{The first decay}

It is now considered a decay of a particle 0 at a given place into two
particles, particle 1 and particle 2. In the decay, there is
conservation of energy, momentum, and electric charge, i.e.,
\begin{equation}
E_0 = E_1 + E_2\,,
\label{eq:energy_penrose}
\end{equation}
\begin{equation}
p_0^r = p_1^r + p_2^r\,,
\label{eq:momentum_penrose}
\end{equation}
\begin{equation}
e_0 = e_1 + e_2\,.
\label{eq:charge_penrose}
\end{equation}
These conservation laws imply directly the following relation,
\begin{equation}
 X_0 = X_1 + X_2\,,
\label{eq:X_penrose}
\end{equation}
see also Appendix \ref{interesting}.

We assume that particle 0
decays with zero velocity
at the inner turning point
$r_{\rm i}$, so 
$\dot r_0=0$. This equation
should have been written as $\dot r_{0\,\rm i}=0$, but
we drop the subscript $i$
to not overcroud the notation, and we will do the same for the
other particles.
From the definition of momentum given in Eq.~\eqref{eq:motion_t_penrose}
and the conservation of momentum given in Eq.~\eqref{eq:momentum_penrose}
one has
$\dot r_0=\dot r_1+\dot r_2$,
where $\dot r_1$ and $\dot r_2$ are the velocities of
particles 1 and 2 at the inner turning
point $r_{\rm i}$, respectively.
Since we assume 
$\dot r_0=0$, we have $0=\dot r_1+\dot r_2$.
We now make a furter assumption, namely, particles 1 and 2 pop out
from the decay with
zero velocity, $\dot r_1=0$ and $\dot r_2=0$, which is a particular
case
consistent with momentum conservation. This simplifies the
calculations and still gives relevant results.  So, we have
\begin{equation}
\dot r_0=0\,,\quad\quad
\dot r_1=0\,,\quad\quad
\dot r_2=0\,.
\label{eq:momentumspeciic_penrose}
\end{equation}
Then, from Eq.~\eqref{eq:motion_t_penrose} we have $P_0 = 0$,
$P_1 = 0$, and $P_2 = 0$.
From Eq.~\eqref{eq:motion_r_penrose} this
implies that
\begin{equation}
X_0
= m_0 \sqrt{f\left(r_{\rm i}\right)}\,,\quad\quad
X_1
= m_1 \sqrt{f\left(r_{\rm i}\right)}\,,\quad\quad
X_2
= m_2 \sqrt{f\left(r_{\rm i}\right)}\,.
\label{eq:Xi_penrose}
\end{equation}
Then, from the conservation law
Eq.~\eqref{eq:X_penrose}, it can be concluded that the mass is
conserved in the decay, i.e.,
\begin{equation}
m_0 = m_1 + m_2\,.
\label{eq:masscons_penrose}
\end{equation}
From Eq.~\eqref{eq:motion_r_penrose}
together with  Eq.~\eqref{eq:Xi_penrose}, the following relations
for the energy of particles 1 and 2 can be obtained,
\begin{equation}
 E_1 = m_1 \sqrt{f\left(r_{\rm i}\right)} +
 \frac{e_1Q}{r_{\rm i}},
    \label{eq:E_1_penrose}
\end{equation}
\begin{equation}
 E_2 = m_2 \sqrt{f\left(r_{\rm i}\right)} +
 \frac{e_2Q}{r_{\rm i}}\,.
    \label{eq:E_2_penrose}
\end{equation}
One can verify then that
Eqs.~\eqref{eq:E_1_penrose} and \eqref{eq:E_2_penrose}
together with the previous equations imply
conservation of energy at the decay,
$E_0=E_1+E_2$,
see Eq.~\eqref{eq:energy_penrose}.

We can now deduce
that for energy extraction to occur,
and so to occur a Penrose process, the energy of
particle 1, which starting with zero velocity
then falls into the black hole, must be negative, i.e., $E_1 <
0$, so that the energy of 
particle 2, which also starting with zero velocity
is then ejected to the outer egion,
is larger than the energy of the decaying one, i.e.,
$E_2> E_0$.
To have 
$E_1 <
0$ means that the decay
must occur inside the electric ergoregion.
The electric ergoregion is defined as the region
in which for a given electric charge
of the particle its energy  is
negative. 
Now, to have $E_1 < 0$ is necessary and sufficient that
first $e_1
= - \lvert e_1 \rvert$, and second, that $\lvert e_1 \rvert
> \frac{r_{\rm i}}{Q} \sqrt{f\left(r_{\rm i}\right)}\, m_1$,
where it was assumed, without loss of generality, that $Q>0$, i.e.,
the black hole has positive electric charge. If the black hole had
negative electric charge, for energy extraction to occur, the electric
charge of odd particles would have to be positive. In what follows, it
will always be considered $Q>0$.  Thus, in brief, for energy
extraction it is needed that the decay happens inside the electric
ergoregion, so that
\begin{equation}
E_{1} <0\,,
\quad\quad\quad
E_{2}> E_{0}\,,
    \label{eq:conditiondecay1initial}
\end{equation}
which implies 
\begin{equation}
\hskip 3.75cm
e_{1} = - \lvert e_{1} \rvert\,,
\quad\quad\quad
\lvert e_{1} \rvert > \gamma m_1\,,\quad\quad
\gamma \equiv\frac{r_{\rm i}}{Q} \sqrt{f\left(r_{\rm i}\right)}\,.
\label{eq:conditiondecay2initial}
\end{equation}
The latter equation gives
the conditions on the electric charge of particle 1.

\subsubsection{Subsequent decays: generic expressions}

After the first decay, i.e., the decay  of particle 0,
it was  considered that particle 1 moves
radially toward the black hole
and particle 2 moves radially outward from the black hole.
Then,
it is supposed that
particle 2 bounces back at a given point, where
the bounce might be due to a reflective mirror
or to a natural outer turning point of
the spacetime.
When particle 2 arrives anew  at
$r_{\rm i}$ it will decay in the same way as particle 0
had done originally, i.e., it will decay
into particle 3 that goes down the hole and into
particle 4 that is ejected out. This
chain of decays will be repeated endlessly,
with the parent particle $2n$ decaying into 
the odd particle $2n+1$ which goes down the hole,
and to the even particle $2n+2$ which is ejected
to be reflected and decay again
at $r_{\rm i}$, with $n=0,1,2,...$. 
We can now generalize the expressions
found above for the first decay to expressions for a generic
subsequent decay.
So here, we consider the
decay of a particle $2n$ at a given place into two
particles, particle $2n+1$ and particle $2n+2$. In the decay, there is
conservation of energy, momentum, and electric charge, i.e.,
\begin{equation}
E_{2n}=E_{2n+1}+E_{2n+2}\,,
\label{eq:consEdecay2n}
\end{equation}
\begin{equation}
p_{2n}^r = p_{2n+1}^r + p_{2n+2}^r\,,
\label{eq:momentum_penrose2n}
\end{equation}
\begin{equation}
e_{2n} = e_{2n+1} + e_{2n+2}\,.
\label{eq:charge_penrose2n}
\end{equation}
These conservation laws imply directly the following relation,
\begin{equation}
 X_{2n} = X_{2n+1} + X_{2n+2}\,.
\label{eq:X_penrose2n}
\end{equation}

We assume that the even particle $2n$
decays with zero velocity
at the inner turning point
$r_{\rm i}$, so 
$\dot r_{2n}=0$. This equation
should have been written as $\dot r_{2n\,\rm i}=0$, but
we drop the subscript i
to not overcroud the notation, and we will do the same for the
other particles.
From the definition of momentum given in Eq.~\eqref{eq:motion_t_penrose}
and the conservation of momentum given in Eq.~\eqref{eq:momentum_penrose}
one has
$\dot r_{2n}=\dot r_{2n+1}+\dot r_{2n+2}$,
where $\dot r_{2n+1}$ and $\dot r_{2n+2}$ are the velocities of
particles ${2n+1}$ and
${2n+2}$ at the inner turning
point $r_{\rm i}$, respectively.
Since we assume 
$\dot r_{2n}=0$, we have $0=\dot r_{2n+1}+\dot r_{2n+2}$.
We now make a further assumption, namely, particles  ${2n+1}$ and
${2n+2}$ pop out
from the decay with
zero velocity, $\dot r_{2n+1}=0$ and $\dot r_{2n+2}=0$,
which is a particular
case
consistent with momentum conservation. Again,
this simplifies the
calculations and still gives relevant results.  So, we have
\begin{equation}
\dot r_{2n}=0\,,\quad\quad
\dot r_{2n+1}=0\,,\quad\quad
\dot r_{2n+2}=0\,.
\label{eq:momentumspeciic_penrose2n}
\end{equation}
Then, from Eq.~\eqref{eq:motion_t_penrose} we have $P_{2n} = 0$,
$P_{2n+1} = 0$, and $P_{2n+2} = 0$.
From Eq.~\eqref{eq:motion_r_penrose} this
implies that
\begin{equation}
X_{2n}
= m_{2n} \sqrt{f\left(r_{\rm i}\right)}\,,\quad\quad
X_{2n+1}
= m_{2n+1} \sqrt{f\left(r_{\rm i}\right)}\,,\quad\quad
X_{2n+2}
= m_{2n+2} \sqrt{f\left(r_{\rm i}\right)}\,.
\label{eq:Xi_penrose2n}
\end{equation}
Then, from the conservation law
Eq.~\eqref{eq:X_penrose2n}, it can be concluded that the mass is
conserved in each decay, i.e.,
\begin{equation}
m_{2n} = m_{2n+1} + m_{2n+2}\,.
\label{eq:masscons_penrose2n}
\end{equation}
From Eq.~\eqref{eq:motion_r_penrose}
together with  Eq.~\eqref{eq:Xi_penrose2n}, the following relations
for the energy of particles ${2n+1}$ and ${2n+2}$ can be obtained,
\begin{equation}
 E_{2n+1} = m_{2n+1} \sqrt{f\left(r_{\rm i}\right)} +
 \frac{e_{2n+1}Q}{r_{\rm i}},
    \label{eq:E_odd_penrose}
\end{equation}
\begin{equation}
 E_{2n+2} = m_{2n+2} \sqrt{f\left(r_{\rm i}\right)} +
 \frac{e_{2n+2}Q}{r_{\rm i}}\,.
    \label{eq:E_even_penrose}
\end{equation}
One can verify then that
Eqs.~\eqref{eq:E_odd_penrose} and \eqref{eq:E_even_penrose}
together with the previous equations imply
conservation of energy at the decay,
$E_{2n}=E_{2n+1}+E_{2n+2}$,
see Eq.~\eqref{eq:consEdecay2n}.

We can now deduce
that for energy extraction to occur,
and so to occur a Penrose process,
the energy of
the odd particle, which
falls into the black hole, must be negative, i.e., $E_{2n+1} <
0$, so that the energy of the
even
particle, which is ejected,
is larger than the energy of the decaying one, i.e.,
$E_{2n+2}> E_{2n}$.
To have 
$E_{2n+1} <
0$ means that the decay
must occur inside the electric ergoregion.
The electric ergoregion is defined as the region
in which for a given electric charge
of the particle its energy  is
negative. 
Now, to have $E_{2n+1} < 0$ is necessary and sufficient that $e_{2n+1}
= - \lvert e_{2n+1} \rvert$, and second, that $\lvert e_{2n+1} \rvert
> \frac{r_{\rm i}}{Q} \sqrt{f\left(r_{\rm i}\right)}\, m_{2n+1}$,
where it was assumed, without loss of generality as mentioned
above, that $Q>0$.
Thus, in brief, for energy
extraction it is needed that the decay happens inside the electric
ergoregion, so that
\begin{equation}
E_{2n+1} <0\,,
\quad\quad\quad
E_{2n+2}> E_{2n}\,,
\label{eq:conditiondecay1}
\end{equation}
which implies 
\begin{equation}
\hskip 3.5cm
e_{2n+1} = - \lvert e_{2n+1} \rvert\,,
\quad\quad\quad
\lvert e_{2n+1} \rvert > \gamma m_{2n+1}\,,
\quad\quad
\gamma \equiv
\frac{r_{\rm i}}{Q} \sqrt{f\left(r_{\rm i}\right)}\,.
\label{eq:conditiondecay2}
\end{equation}
The latter equation gives
the conditions on the electric charge of particle $2n+1$.

With these results we can now introduce two different scenarios for
the decay of the particles, scenario 1 and scenario 2. The most
important difference between the two scenarios, whose specific
conditions will be indicated in the following section, is that in the
first scenario the electric charge of the particles is fine tuned to
allow energy extraction in every decay, starting from the first one,
while in the second scenario energy extraction can start occurring
after a few numbers of decays. See also \cite{penrosekok,penrosezasl}.
These two scenarios lead to black hole energy factories and possibly
to black hole bombs.

\subsection{The two different scenarios: Scenario 1 and scenario 2}
\label{sec:twoscenarios}

\subsubsection{Scenarios 1 and 2: Mass decay and mass
conservation}

In both scenarios, 1 and 2, one can write the masses
of the decaying particles $m_{2n+1}$ and  $m_{2n+2}$
in terms of the mass $m_{2n}$ as
\begin{equation}
 m_{2n+1}=\alpha_1 m_{2n}\,,
    \label{eq:mass_odd_case1_penrose}
\end{equation}
\begin{equation}
 m_{2n+2} =\alpha_2 m_{2n}\,,
    \label{eq:mass_odd_case1_penrose2}
\end{equation}
for some parameters $\alpha_1$ and $\alpha_2$
that are fixed, i.e., have the same value for all
the decays, and
due to mass conservation in the decay one has then
\begin{equation}
\alpha_1 + \alpha_2 = 1\,,
\label{eq:massconsalpha}
\end{equation}
see also Eq.~\eqref{eq:masscons_penrose}.
Since the masses are positive, one has $\alpha_1>0$ and $\alpha_2>0$.
The two scenarios differ in
the electric charges
of the decayed particles, see below.

The conditions for the masses of odd and even particles,
Eqs.~\eqref{eq:mass_odd_case1_penrose}
and \eqref{eq:mass_odd_case1_penrose2},
can be solved iteratively, 
leading to the general expressions
\begin{equation}
m_{2n+1} =\alpha_1 (1-\alpha_1)^n m_{0}\,,
\label{eq:mass2_case1_penrose}
\end{equation}
\begin{equation}
m_{2n+2}=\alpha_2^{n+1} \,m_{0}\,,\quad\quad\quad
\label{eq:mass1_case1_penrose}
\end{equation}
where use of Eq.~\eqref{eq:massconsalpha} was also made.
Note that as the number of decays
tends to infinity, $n\to\infty$, one has that
the decaying masses behave as
$m_{2n+1}\to0$ and $m_{2n+2}\to0$.

\subsubsection{Scenarios 1 and 2: Charge decay and charge
conservation}

In scenario 1, it is considered that all odd particles,
starting from the first decay, have a negative electric charge, lower
than $ \gamma m_{2n+1}$. This ensures that the energy of odd particles,
given by Eq.~\eqref{eq:E_odd_penrose}, is always negative, and
therefore the energy of even particles is larger than the energy of
the previous decaying particle, during the whole chain of
decays. Then, it is straightforward to see that, in this scenario,
the Penrose process leads
to energy extraction from the black hole.
The conditions of charge deployment assumed for this case can be
expressed as
\begin{equation}
 e_{2n+1}= -\left(1+\chi\right)  \gamma m_{2n+1}\,,
    \label{eq:charge_odd_case1_penrose}
\end{equation}
\begin{equation}
 e_{2n+2}= e_{2n} - e_{2n+1}\,,
    \label{eq:charge_even_case1_penrose}
\end{equation}
where $\chi$ is some number greater than zero. Clearly, 
this
deployment of the charge satisfies the charge
conservation law
\begin{equation}
-\left(1+\chi\right)  \gamma m_{2n+1}
+ e_{2n+2}= e_{2n}\,,
\label{eq:chargeconsscenario1}
\end{equation}
see also Eq.~\eqref{eq:charge_penrose}, 
and it implies a negative energy for odd particles in
all decays, starting from the first one. This
scenario leads to  black hole
energy factories.

In scenario 2,
the deployment of the electric charges is different
and is such
that the energy of the odd particle does not need to be negative from
the first decay, but it decreases with $n$, becoming negative after
some value of $n$, and therefore allowing energy extraction
with the Penrose process as well. This assumption can be established in
terms of the following charge deployment
\begin{equation}
\hskip 2.2cm   e_{2n+1}= \beta_1 e_{2n}\,,\quad\quad  \beta_1<0\,,
\label{eq:charge_odd_case2_penrose}
\end{equation}
\begin{equation}
\hskip 2.2cm  e_{2n+2}= \beta_2 e_{2n}\,,\quad\quad  \beta_2>1\,,
\label{eq:charge_even_case2_penrose}
\end{equation}
for some parameters $\beta_1$ and $\beta_2$
that are fixed, i.e., do not depend on
the number of decays, and due
to charge conservation in the decay one has then
\begin{equation}
\beta_1 + \beta_2 = 1\,,
\label{eq:chargeconsbeta}
\end{equation}
see also Eq.~\eqref{eq:charge_penrose}.
As we see,
it is also assumed that $\beta_1 < 0$ and, therefore, $\beta_2 > 1$.
This
scenario may lead to  black hole bombs.

\subsection{Confinement situations with mirror and with no mirror, and
scenarios 1 and 2}

Thus, in Reissner-Nordstr\"om-AdS black hole spacetimes there are two
situations to realize a recursive Penrose process.
One situation has a reflective spherical mirror at
some radius $r_{\rm m}$. The other situation has no mirror.

Within each situation there are two possible scenarios for the
decayment of the particles, scenario 1 and scenario 2.  In both
scenarios the masses of the particles involved obey the relations
given in Eqs.~\eqref{eq:mass_odd_case1_penrose} and
\eqref{eq:mass_odd_case1_penrose2}.  The difference in the scenarios
appears in the charge deployment.  In scenario 1, the ratio
$\frac{e_{2n+1}}{m_{2n+1}}$ remains constant, see
Eqs.~\eqref{eq:charge_odd_case1_penrose} and
\eqref{eq:charge_even_case1_penrose}.  In scenario 2 the ratios
$\frac{e_{2n+1}}{e_{2n}}$ and $\frac{e_{2n+2}}{e_{2n}}$ are equal to
two different constants, see Eqs.~\eqref{eq:charge_odd_case2_penrose}
and \eqref{eq:charge_even_case2_penrose}.
\vskip 0.1cm

So,
we first discuss Reissner-Nordstr\"om-AdS black hole spacetimes with a
reflective mirror at $r_{\rm m}$ in scenario 1 and scenario 2 and
study the possibility of black hole energy factories and
bombs, and second we discuss pure Reissner-Nordstr\"om-AdS black hole
spacetimes, i.e., with no mirror, in scenario 1 and scenario 2 and study the
possibility of black hole energy factories and bombs. This
second possibility can be considered as the limit of the first
possibility when $r_{\rm m}$ is sent to infinity.

\section{Penrose process in Reissner-Nordstr\"om-AdS black
hole spacetimes in the situation
with a reflective mirror: Black hole energy
factories and black hole bombs}
\label{sec:confined}

\subsection{Scenario 1: Energy extraction and black hole
energy factory}
\label{sec:scenario1}

\subsubsection{Expressions for the
electric charge and the energy after $n+1$ decays}

The conditions for the masses of odd and even particles,
yield
$m_{2n+1} =\alpha_1 (1-\alpha_1)^n m_{0}$  and
$m_{2n+2}=\alpha_2^{n+1} \,m_{0}$, as we have seen in
Eqs.~\eqref{eq:mass2_case1_penrose} and 
\eqref{eq:mass1_case1_penrose}, respectively.

For scenario 1,
the conditions for the electric charge of odd and even particles,
Eqs.~\eqref{eq:charge_odd_case1_penrose}
and
\eqref{eq:charge_even_case1_penrose}, can also be solved
in this scenario. For the odd
particles one finds
\begin{equation}
e_{2n+1} = - 
\left(1+\chi\right)\gamma\, \alpha_1 (1-\alpha_1)^n\,m_0.
\label{eq:charge1_case1_penrose}
\end{equation}
For the even particles,
full expressions can also be obtained, assuming that
the particles are reflected in the mirror,
and bounce back to the decaying point, yielding
\begin{equation}
e_{2n+2} = e_0 + \left(1+\chi\right)\gamma
\left(1-\alpha_2^{n+1}\right)\, m_0,
\label{eq:charge_2n_case1_penrose}
\end{equation}
Note that as $n\to\infty$ one has
$e_{2n+1}\to0$ and $e_{2n+2}\to e_0 + \left(1+\chi\right)\gamma \,m_0$.

Substituting the previous results into Eqs.~\eqref{eq:E_odd_penrose}
and \eqref{eq:E_even_penrose},
the energies of odd and even
particles after $n+1$ decays can be obtained
as
\begin{equation}
E_{2n+1} = - \sqrt{f\left(r_{\rm i}\right)}\, \chi\,m_{2n+1}\,,
\label{eq:energyn1_case1_penrose2}
\end{equation}
\begin{equation}
E_{2n+2} = E_0 +  \chi\gamma
\left(1-\alpha_2^{n+1}\right) \frac{Q}{r_{\rm i}}\, m_0\,,
\label{eq:energy_2n_case1_penrose}
\end{equation}
respectively.

\subsubsection{Analysis of the turning points in scenario 1}

To understand the effect of the reflective mirror in a
Reissner-Nordstr\"om-AdS black hole spacetime, it is necessary to
study the positions of the turning points for the different particles
participating in the decay chain. The turning points can be obtained
from Eq.~\eqref{eq:turning_points_penrose}. There is an inner turning
point $r_{\rm i}$, where the decay is assumed to happen and an outer
turning point $r_{\rm t}$, corresponding to the largest zero of
Eq.~\eqref{eq:turning_points_penrose}.
As Eq.~\eqref{eq:turning_points_penrose} only depends on the energy
per unit mass of the particles, i.e., $\frac{E}{m}$, and on the charge
per unit mass, i.e., $\frac{e}{m}$ of the particles, the position of
the largest turning point can only depend on these two
quantities. This outer turning point $r_{\rm t}$ is to be evaluated
for even particles since, as shown in Appendix \ref{sec:odd_dynamics},
the odd particles fall into the black hole. It can be seen from
Eq.~\eqref{eq:energyn1_case1_penrose2} that
$\frac{E_{2n+1}}{m_{2n+1}}$ does not depend on $n$. It follows from
Eqs.~\eqref{eq:mass2_case1_penrose}
and~\eqref{eq:charge1_case1_penrose} that $\frac{e_{2n+1}}{m_{2n+1}}$
does not depend on $n$ but from Eq.~\eqref{eq:charge_2n_case1_penrose}
$\frac{e_{2n+2}}{m_{2n+2}}$ does.  It can also be proved that, in this
decaying scenario, the outer turning point has a finite position after
an infinite number of decays, see Appendix \ref{sec:turning_position}.


Now, there are two options of placing the reflective spherical
mirror. If the mirror is placed in a radius $r_{\rm m}$ lower than the
initial position of the outer turning point $r_{\rm t}
\left(n=0\right)$, i.e., $r_{\rm m} < r_{\rm t} \left(n=0\right)$, the
outer turning point is always unreachable for the even particles.
Thus, even particles will always be reflected by the mirror, bouncing back
to the decaying point, in the same way as in Reissner-Nordstr\"om
black hole spacetime \cite{penrosekok,penrosezasl}.  If the reflective
mirror is placed outside the outer turning point, i.e., $r_{\rm m} >
r_{\rm t} \left(n=0\right)$, the mirror is unreachable if $r_{\rm m} >
r_{\rm t} \left(n=\infty\right)$ or, the first option is recovered if
$r_{\rm m} < r_{\rm t} \left(n=\infty\right)$.  In the case with an
unreachable mirror, the turning point  acts in the same way
as the reflective mirror, since the even particles bounce back from
the turning point into the black hole. Therefore, in the
Reissner-Nordstr\"om-AdS black hole spacetime a decaying chain is
also possible in this case.

\subsubsection{The black hole energy factory}

In the previous section, it was shown that a sequence of $n+1$ decays
can occur with
the occurrence of energy extraction from the black
hole in each decay. Now,
$n$ can be
arbitrarily large. After an infinite number of decays,
$n\to \infty$, 
since
$\alpha_2 < 1$, the energy remains
finite and is given by
\begin{equation}
 \lim_{n\to\infty} E_{2n+2} = E_0 + 
 \chi\gamma \frac{Q}{r_{\rm i}}\, m_0 .
    \label{eq:energy_infinity_case1_penrose}
\end{equation}
Therefore, the efficiency $\eta_n \equiv
\frac{E_{2n+2}}{E_0}$ is also finite, meaning that black hole bomb is
not possible in this decaying scenario, in the same way as 
for Reissner-Nordstr\"om spacetime \cite{penrosezasl}.

It can also be seen from Eq.~\eqref{eq:energy_infinity_case1_penrose}
that the energy $E_{2n+2}$ after an infinite number of decays is
larger than the energy of the initial particle, $E_0$. Thus, there is
an energy gain due to the recursive Penrose process.  Note that if the
reflective mirror is placed before the final position of the turning
point, i.e., $r_{\rm m} < r_{\rm t}\left(n=\infty\right)$, the
particles are confined to a finite
volume $V\left(r_{\rm m}\right)$
of radius $r_{\rm m}$.  In turn, if
the
mirror is placed outside the turning point, i.e.,
$r_{\rm m}>r_{\rm t}\left(n=\infty\right)$, the motion is confined to a
finite volume $V\left(r_{\rm t}\right)$
of radius
$r_{\rm t}$, see Appendix~\ref{sec:volume} for
the calculation of the volumes. In both cases there is
an energy gain confined to 
finite space regions.
Therefore, in each case, the energy per unit volume stored
in the box inside the mirror
after an
infinite number of particle decays is given by
\begin{equation}
\lim_{n\to\infty} \frac{E_{2n+2}}{V\left(r_{\rm m}\right)} = \frac{E_0
+ \chi\gamma \frac{Q}{r_{\rm i}}\,m_0}{V\left(r_{\rm
m}\right)}\,,\quad\quad\quad \lim_{n\to\infty}
\frac{E_{2n+2}}{V\left(r_{\rm t}\right)} = \frac{E_0 + \chi\gamma
\frac{Q}{r_{\rm i}}\,m_0}{V\left(r_{\rm
t}\left(n=\infty\right)\right)}\,.
\label{eq:energy_infinity_case1_penrose2}
\end{equation}
These
energy densities are positive and finite, which means
that a finite energy gain is
stored in a finite region, i.e., in a box,  of the space.
Thus, this leads to  black hole energy factories.

\subsection{Scenario 2: Energy extraction and black hole bomb}
\label{sec:scenario2}

\subsubsection{Expressions for the
electric charge and the energy after $n+1$ decays}

The conditions for the masses of odd and even particles,
yield
$m_{2n+1} =\alpha_1 (1-\alpha_1)^n m_{0}$  and
$m_{2n+2}=\alpha_2^{n+1} \,m_{0}$, as we have seen in
Eqs.~\eqref{eq:mass2_case1_penrose} and 
\eqref{eq:mass1_case1_penrose}, respectively.

For scenario 2, the following relations for the
electric charges can be obtained from
Eqs.~\eqref{eq:charge_odd_case2_penrose} and
\eqref{eq:charge_even_case2_penrose},
\begin{equation}
 e_{2n+1}= \beta_1 (1-\beta_1)^n\, e_{0}\,,
    \label{eq:charge2_case2_penrose}
\end{equation}
\begin{equation}
 e_{2n+2}= \beta_2^{n+1} e_{0}\,.
    \label{eq:charge1_case2_penrose}
\end{equation}
Note that as $n\to\infty$ one has
$e_{2n+1}\to - \infty$ since
$\beta_1<0$,
and $e_{2n+2}\to \infty$ since
$\beta_2>1$.


Using
Eqs.~\eqref{eq:E_odd_penrose} and \eqref{eq:E_even_penrose}, the
expressions for the energy of odd and
even particles after $n+1$ decays
can be
obtained as
\begin{equation}
 E_{2n+1} =  \left(\sqrt{f\left(r_{\rm i}\right)} -
 \frac{e_0 Q}{r_{\rm i} m_0}
 \frac{\lvert\beta_1\rvert}{\alpha_1}
 \left(\frac{1-\beta_1}{1-\alpha_1}\right)^n\right)\, m_{2n+1}\,,
    \label{eq:energy1_case2_penrose}
\end{equation}
\begin{equation}
 E_{2n+2} = \left(\sqrt{f\left(r_{\rm i}\right)} +
 \frac{e_0 Q}{r_{\rm i} m_0}\left(\frac{\beta_2}{\alpha_2}\right)^{n+1}
 \right)\,m_{2n+2}\,,
    \label{eq:energy2_case2_penrose}
\end{equation}
respectively,
where it was assumed that particles are able to bounce back from the
reflective mirror to the decaying point.

\subsubsection{Analysis of the turning points in scenario 2}
\label{sec:scenario2turning}

There is one inner turning point where the
decays are assumed to occur and an outer turning point, whose position
depends on $\frac{E}{m}$ and  $\frac{e}{m}$ of the particles. From 
Eqs.~\eqref{eq:mass2_case1_penrose},
\eqref{eq:mass1_case1_penrose},  and
\eqref{eq:charge1_case2_penrose}, the electric
charge of the even
particles can be computed as
\begin{equation}
e_{2n+2} = \left(\frac{\beta_2}{\alpha_2}\right)^{n+1}\,
\frac{m_{2n+2}}{m_0}\, e_0 .
\label{eq:charge2_case2_penrose2}
\end{equation}
Since $\beta_2 > 1$ and $\alpha_2 < 1$, the charge per unit mass of
even particles increases with $n$, and therefore the outer turning
point changes its position.  From
Eq.~\eqref{eq:energy2_case2_penrose}, the energy of the even particles
can be written as $E_{2n+2} = \left(\sqrt{f\left(r_{\rm i}\right)} +
\frac{e_0 Q}{r_{\rm i} m_0}\left(\frac{\beta_2}{\alpha_2}\right)^{n+1}
\right)\,m_{2n+2}$, which in turn can be put in the form
\begin{equation}
E_{2n+2} = \left(\sqrt{f\left(r_{\rm i}\right)} +
\frac{Q}{r_{\rm i}} \frac{e_{2n+2}}{m_{2n+2}}\right)\,m_{2n+2}
\,.
\label{eq:epsilon_case2_penrose}
\end{equation}
Thus, both charge and energy of the even
particles increase with $n$.

Again, the outer turning point $r_{\rm t}$ is to be evaluated for even
particles only since, as shown in Appendix \ref{sec:odd_dynamics}, the
odd particles fall into the black hole.
It can also be shown, see
Appendix \ref{sec:turning_position},
that in scenario 2
the outer turning point, after a large number of
decays, and therefore for large values of $\frac{E}{m}$ and
$\frac{e}{m}$, grows with the square root of $\frac{E}{m}$, which means
that the outer turning point moves away from the black hole along the
decay chain, reaching infinity after an infinite number of decays.
Thus, if a reflective mirror is placed
at some radius $r_{\rm m} > r_{\rm t}\left(n=0\right)$,
the emitted particle of
the first decay will bounce back to the decaying point from the outer
turning point at $r_{\rm t}\left(n=0\right)$. However, after a certain
number $n_{\rm n}$
of decays,  since the turning point moves further away
from the black hole, the reflective mirror will be before the turning
point, i.e., $r_{\rm m} < r_{\rm t}\left(n=n_{\rm n}\right)$.
Then, emitted particles
will bounce back from the reflective mirror, not reaching the outer
turning point. This will guarantee the existence of a decaying chain
up
to an infinite number of decays.

\subsubsection{The black hole bomb}

The decay chain in this spacetime gives that 
as $n \to \infty$, one has that $E_{2n+2}$ diverges, as can
be seen from  Eqs.~\eqref{eq:charge2_case2_penrose2}
and
\eqref{eq:epsilon_case2_penrose}. Thus,
\begin{equation}
\lim_{n\to\infty} E_{2n+2} =   \infty\,.
\label{eq:energy_infinity_case2_penrose}
\end{equation}
Indeed, $E_{2n+2}$ grows exponentially with $n$.
As well, the efficiency $\eta_{n}\equiv
\frac{E_{2n+2}}{E_0}$ is given by $\lim_{n\to\infty} \eta_{n} = + \infty$.

After an infinite number of decays,
the outer turning point moves to infinity, but the particles remain
confined between the inner turning point and the reflective mirror,
placed at $r_{\rm m}$. Therefore,  an
infinite energy gain is confined to a finite space region of
finite volume $V\left(r_{\rm m}\right)$,
see Appendix \ref{sec:volume} for
the complete calculation of the volume. Thus, the energy per unit
volume diverges,
\begin{equation}
\lim_{n\to\infty} \frac{E_{2n+2}}{V\left(r_{\rm m}\right)} =  \infty,
\label{eq:energy_infinity_case2_penrose2}
\end{equation}
which means that there is a black hole bomb in this scenario.

\subsection{Synopsis of both scenarios}

In the situation where a
reflective mirror is present,
two scenarios of decay have been analyzed.
In the first scenario, after an infinite
number of decays, the recursive Penrose process
yields
an energy gain that is finite. This configures a black
hole energy factory, in which a finite energy extracted from the
black hole is confined to a finite volume.  In the second scenario, it
was shown that  after an infinite
number of decays there is an infinite gain of energy confined in a
finite volume, and so this scenario allows a black hole bomb.

\section{Penrose process in Reissner-Nordstr\"om-AdS
spacetime in the situation with no mirror: Black hole
energy factory and demined black hole bomb}
\label{sec:no_mirror}

\subsection{Scenario 1: Energy extraction and black hole
energy factory}

In scenario 1, the energy
after $n$ decays for odd and even particles is given by
Eq.~\eqref{eq:energyn1_case1_penrose2} and
\eqref{eq:energy_2n_case1_penrose}. As it was shown in
Sec.~\ref{sec:scenario1}, the outer turning point, in this 
scenario, has a fixed position during the whole decay
chain. Therefore, in the absence of a reflective mirror
which is the situation we are considering now, even particles
bounce back from the outer turning point, reaching the decaying position
again, allowing a decay chain to happen. This means that the outer
turning point can exactly replace the effect of a reflective mirror in
the decaying chain. Thus, after an infinite number of decays,
there will be an energy gain, given by the same amount as in the
presence of a reflective mirror,
$
 \lim_{n\to\infty} E_{2n+2} = E_0 + 
 \chi \gamma \frac{Q}{r_{\rm i}}\,m_0$.
In fact, this scenario has the same expression
for the energy gain as the one in which the
reflective mirror is placed outside the turning point, i.e., $r_{\rm
m} > r_{\rm t}\left(n=\infty\right)$,
see Eq.~\eqref{eq:energy_infinity_case1_penrose}.
The finite energy gain is
confined to a finite volume in spacetime. However, this volume is now
not bounded by a reflective mirror, and therefore there is no
physical box
where the energy is confined. However, since particles cannot move to
radii larger than $r_{\rm t}\left(n=\infty\right)$, the energy is
confined in a region of Reissner-Nordstr\"om-AdS spacetime, so that
the spacetime functions as a black hole energy factory.

\subsection{Scenario 2: Energy extraction and demined
black hole bomb, i.e., no bomb}

Assuming that a decaying chain is possible for this scenario without a
reflective mirror, the energy of odd and even particles is given by
Eqs.~\eqref{eq:energy1_case2_penrose} and
\eqref{eq:energy2_case2_penrose}.  As shown in
Sec.~\ref{sec:scenario2}, the position of the outer turning point
changes along the decay chain, since it depends on
$\frac{E_{2n+2}}{m_{2n+2}}$ and $\frac{e_{2n+2}}{m_{2n+2}}$, quantities
which depend on $n$. Therefore, it is necessary to verify if the the
outer turning point remains outside the ergoregion and if the motion
of particles is possible until reaching the inner turning point,
before crossing the horizon.

\begin{figure}[h]
    \centering
    \includegraphics[width = 0.47\textwidth, height =
    0.2\textheight]{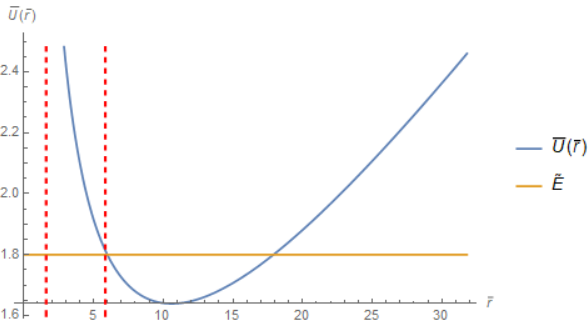}
\caption{It is plotted a rescaled potential energy, as a function of
the rescaled radius for the electrically charged particle
in a Reissner-Nordstr\"om-AdS black hole spacetime.
The rescaled quantities are $\bar
r=\frac{r}{M}$, $\bar Q=\frac{Q}{M}$, $\bar l =\frac{l}{M}$, $\tilde
E=\frac{E}{m}$, $\tilde e=\frac{e}{m}$, $f(\bar r) =
\frac{\frac{{\bar r}^4}{{\bar l}^2} + {\bar r}^2 - 2{\bar r}
+{\bar Q}^2}{{\bar r}^2}$, and ${\bar U}(\bar r)=\sqrt{f(\bar r)}+
\frac{\tilde{e} \bar{Q}}{\bar r}$.  The values used are
$\bar{Q}=0.78$, $\bar{l}=15.3$, $\tilde{e} = 6.8$, and $\tilde{E} =
1.8$. Note that ${\bar E}\geq {\bar U}$ in this plot, as it is the
case when motion is allowed.  One vertical line shows the 
turning point ${\bar r}_0$,which was chosen to be
at the limit of the
ergoregion, so that ${\bar r}_0=5.8$, and the
other vertical line shows the event horizon radius ${\bar r}_+=1.6$.
}
\label{fig:g_x}
\end{figure}

To study where motion is allowed
we find that motion is
only possible for radii which satisfy $X^2 - m^2 f\left(r\right) \geq
0$, since the square root in Eq.~\eqref{eq:motion_r_penrose} has to
be non-negative. This condition can be rewritten in terms of the
energy as
\begin{equation}
E\geq U(r)  \,,
    \label{eq:condition_epsilon}
\end{equation}
where the potential energy $U(r)$ is defined as
\begin{equation}
U(r)=m\sqrt{f\left(r\right)}+\frac{e Q}{r}   \,.
    \label{eq:condition_epsilon2}
\end{equation}
We see that
motion is possible for radii where the energy  of the
particle is larger than the effective potential energy
$U(r)$. In Fig.~\ref{fig:g_x},
it is shown the effective potential for specific
values of black hole's charge, cosmological constant, particle's
specific charge, and particle's energy, all quantities
suitably rescaled.

The effect of the decaying chain is to increase both particle's electric
charge and energy. For the decaying chain to continue up to $n \to
\infty$, the particle must remain confined between two turning points,
one inside or at
the ergoregion and one outside. The inner turning point is
not significantly changed by the presence of the cosmological
constant, since it happens for low radii, therefore the term
proportional to the cosmological constant can be neglected and it is
assumed that all decays happen in the same position, corresponding to
the inner turning point. The presence of the outer turning point
has to be guaranteed in order to allow the sequence of
decays in the absence of a reflective mirror.

When $n$ increases, the electric charge of
the even particles
$e_{2n+2}$ increases. This will contribute to an increase of the
minimum of $U(r)$, i.e.,
an increase of $U_{\rm min}$. So $U_{\rm min}$ is a
growing function of $e$. However,
the even particle's energy
also increases with $n$, indeed the particle's energy
per
unit mass increases with $n$.
From Eq.~\eqref{eq:epsilon_case2_penrose}
we see that the increase in energy  is related
to the increase in the electric charge of the particle..
Therefore, if in the limit of
$e \to \infty$, $E(e)$ remains
larger than $U_{\rm min}(e)$, then the  existence of the
outer turning point is guaranteed, and even particles will have a
trapped motion between
the two turning points, as needed for the sequence
of decays.
Let us see this in detail.
In the limit of large $n$, one has
$e_{2n+2} \to \infty$, and the effective
potential energy $U(r)$
of Eq.~\eqref{eq:condition_epsilon2}
can be simplified into
$\lim_{n\to\infty}U(r)=m\frac{r}{l}+\frac{e Q}{r}$,
where the first term cannot be neglected, since it is the dominant term
for large values of radius $r$. This
function
has a minimum at
$r_{\rm min} = \sqrt{\frac{e}{m}Q \,l}$.
The minimum of $U\left(r\right)$, as
$e \to \infty$ is, therefore, given by
\begin{equation}
U_{\rm min}\left(e\right) = 2
\sqrt{\frac{Q}{l}\frac{e}{m}}\,,
\label{eq:potential_large_e2}
\end{equation}
so that $U_{\rm min}(e)$ goes with $\sqrt{\frac{e}{m}}$
for large $n$.
Thus, the minimum of $U(r)$ scales with the square root of
particle's charge per unit mass when $n \to \infty$. On the other
hand, the even particle's energy, according to
Eq.~\eqref{eq:epsilon_case2_penrose}, for large values of the electric
charge, scales linearly with it. Therefore, for large
values of $n$, the energy increases faster than the
minimum of the effective potential energy,
and so 
the energy per unit mass increases faster than the
minimum of the effective potential per unit mass.
This guarantees the presence of
the two needed turning points for the chain of decays, implying that,
after an infinite number of decays,
from Eq.~\eqref{eq:energy2_case2_penrose},
the energy of even particles
diverges as ${\rm e}^{n\ln\frac{\beta_2}{\alpha_2}}$, i.e.,
$E_{2n+2}$ grows exponentially with $n$, so that
\begin{equation}
\lim_{n\to\infty} E_{2n+2} = \infty\,.
\label{eq:energy_infinity_case2_penrose_no_mirror}
\end{equation}
The efficiency then obeys
$ \lim_{n\to\infty} \eta_{n} =  \infty$.
Does it mean that Reissner-Nordstr\"om-AdS is a
potential black hole bomb?
To answer the question
note
that the outer turning point changes its position during the
chain of decays,
and the volume where the energy diverges also changes
with $n$. Thus, to verify
if a black hole bomb is possible in the
absence of a reflective mirror, it is necessary to
verify whether the energy
per unit volume diverges or not.
If not, it would be interesting
to know if it remains finite when $n \to \infty$.

The position of the outer turning point $r_{\rm t}$ as a function of the
particle and spacetime parameters, can be obtained from
Eq.~\eqref{eq:turning_points_penrose}. In the limit of $n \to \infty$, it
can be shown that the outer turning point position scales as
$\sqrt{\frac{E_{2n+2}}{m_{2n+2}}}$,
see Appendix \ref{sec:turning_position}.
Therefore, the particles are confined to a growing volume of radius
$r_{\rm t}$, that scales with
$\sqrt{E_{2n+2}}$, and so
increasing with $n$. It is
shown in the Appendix \ref{sec:volume} that, for large values of $n$, the
volume $V(r_{\rm t})$ is proportional to $r_{\rm t}^2$.
Thus, this implies that
$\lim_{n\to\infty} \frac{E_{2n+2}}{r_{\rm t}^2} = \frac{m_{2n+2}\;
E_{2n+2}}{E_{2n+2}} \to 0$,
and since $m_{2n+2}\to 0$ when $n\to\infty$,
we have
\begin{equation}
\lim_{n\to\infty} \frac{E_{2n+2}}{V(r_{\rm t})} \to 0\,.
\label{eq:no_BH_bomb}
\end{equation}
Therefore, although the energy diverges when $n \to \infty$, there is
no black hole bomb, since the volume where the particles are confined
grows faster with $n$ than the energy and so the energy per unit
volume decreases along the decay chain. The boundary of the volume
where the energy gain is confined is being pushed to infinity along
the decay chain, demining the black hole bomb to a no bomb limit, when
the boundary is at infinity. This situation configures
thus a demined black
hole bomb. The same situation can be obtained from considering the
reflective mirror introduced in Sec.~\ref{sec:scenario2} placed now at
infinity. In this situation, the outer turning point is always before
the mirror, i.e., $r_{\rm m} > r_{\rm t}$, turning it unreachable.
Note that here there is no  energy factory
as well.

\subsection{Synopsis of both scenarios}

In the situation where a
reflective mirror is absent,
i.e., in pure Reissner-Nordstr\"om-AdS
black hole spacetime, 
the  negative cosmological constant
acts like a mirror and
introduces a new
turning point outside the ergoregion.
Thus, it is possible that this
outer turning point can be used to replace the reflective mirror.
The same two scenarios of decay were analyzed
for pure Reissner-Nordstr\"om-AdS black hole spacetime.
In the first scenario, it was shown that the outer turning point has a
finite position during the whole decaying chain, acting 
as an effective
reflective mirror, thus
allowing for a recursive Penrose process. The
energy extracted from the black hole in this situation
has the same expression as the one
with a reflective mirror.
Although, here there is no
physical boundary,
particles still remain confined in a finite volume region,
due to the presence of the turning point.
Thus, this situation characterizes also a black hole energy
factory.
In the second scenario, it was shown that the new outer turning point
introduced by the negative cosmological constant allows
for a confined
motion of the particles.
It was found that the position
of this turning point
depends on the ratio
between the electric charge and the mass of the particle and on the
ratio between the energy of the particle
and its mass. Therefore, contrary to the reflective mirror,
during the whole chain of
decays, the turning point
does not have a fixed
position.
We found that the energy of the
outgoing particles diverges after an infinite number of decays.
However, the position of the turning point was shown to increase
with energy, which means that the volume where particles are
confined increases along the decaying chain. It
increases in such a way that the energy per unit volume
goes to zero,
and therefore a black hole bomb is not possible in this
scenario. The bomb is demined as the
boundary of the volume is pushed to infinity and, in fact,
there is not any energy factory.
In Table \ref{table}, it is summarized the behavior of mass, charge,
energy, and outer turning point 
of even and odd particles after an infinite number of
decays, i.e., when $n \to \infty$
for the no mirror situation in both scenarios.
\begin{table}[h]
\begin{tabular}{ c| c| c }
 $\lim_{n\to \infty}$ & Scenario 1 & Scenario 2 \\ \hline
 $m_{2n+1}$ & 0 & 0 \\  
 $m_{2n+2}$ & 0 & 0 \\
 $e_{2n+1}$ & 0 & $-\infty$ \\  
 $e_{2n+2}$ & $e_0 + \left(1+\chi\right) \gamma \, m_0$ & $+\infty$ \\
 $E_{2n+1}$ & 0 & $-\infty$ \\  
 $E_{2n+2}$ & $E_0 + \chi \gamma \frac{Q}{r_{\rm i}} m_0$ & $+\infty$  \\
  $r_{\rm t}$ & finite value & \;$ \propto\sqrt{\frac{E}{m}} \to + \infty$   
\end{tabular}
\caption{Behavior of mass, charge, energy and outer turning point of
particles when $n \to \infty$
in a recursive Penrose process
in a Reissner-Nordstr\"om black hole spacetime
in the no mirror situation in scenarios 1 and 2.}
\label{table}
\end{table}

\section{Conclusions}
\label{sec:concl}

We have analyzed the recursive Penrose process
for
electrically charged particles in a Reissner-Nordstr\"om-AdS black
hole spacetime
in two situations, when
there is a mirror and when there is none, and for each
situation we have applied two scenarios for the decays.
With a mirror we have shown that black hole energy
factories and black hole bombs are possible, and without a mirror,
i.e., pure Reissner-Nordstr\"om-AdS black hole
spacetime, we have shown that black hole
energy factories exist, but there is no black hole bomb.

For both situations, with mirror and without mirror in a
Reissner-Nordstr\"om-AdS black hole spacetime,
and for both scenarios, 1 and 2, one can ask what is the
end product of the system.  Our calculations have been done for
fixed background and so say nothing about this point,
nevertheless one can explore the possibilities.
In scenario 1, the electric charge extraction and energy gain
are both finite, and so the backreaction is negligible,
or almost negligible, meaning
that the initial black hole does not change much its initial
mass $M$ and charge $Q$ after an infinite number of decays.
In scenario 2 it is different, since
the electric charge extraction and energy gain go to infinite.
At some stage in the recursive
process the
the energy of the particle would be of the order of the
black hole mass
$M$ and the
electric charge of the particle would be of
the order of the black hole charge $Q$, and surely this cannot be.
This implies that backreaction, not only on the spacetime metric but
also on the electric charge cannot be ignored on physical grounds.
When there is a mirror and so a bomb, in this scenario 2, one can
predict that the Penrose process conditions fail to be met after some
large number $n$ of decays, and the last even particle cannot be
ejected outward anymore.  The Penrose process itself stops from then
on. One is left with a black hole of lower mass
$M$ and lower electric charge $Q$
than the initial values, and with a particle hanging at
some definite radius. This characterizes a hairy black hole in much
the same way of what has been found in superradiance, where
electrically charged scalar fields in Reissner-Nordstr\"om-AdS black
holes have a linear instability that evolves nonlinearly into a hairy
black hole.  So it seems that the recursive Penrose process also yields,
taking into account backreaction, 
a hairy black hole, the hair being constituted of a single particle
counterpoised at some radius. Clearly, backreaction
demines the black hole bomb.
When there is no mirror and so the bomb is demined
even without backreaction, in this scenario
2, one can predict again that the Penrose process conditions fail to
be met after some large number $n$ of decays, the even particle cannot
be ejected outward, and the Penrose process itself stops from then on.
The end state would be again a hairy black hole with the electrically
charged particle counting as hair.  Clearly, backreaction demines the
black hole bomb at a finite $n$, before it is demined for $n\to\infty$
in the analysis without backreaction.  These latter comments are
qualitative in character. To obtain a concrete result, one must
include the black hole mass and charge losses at each Penrose process
step in view of backreaction.

\acknowledgments{We acknowledge financial support from Funda\c c\~ao
para a Ci\^encia e Tecnologia - FCT through the
project~No.~UIDB/\break 00099/2020.}

\appendix

\section{The horizon radii of the Reissner-Nordstr\"om-AdS spacetime}
\label{hradii}

From Eq.~\eqref{eq:f_RN_AdS},
we have that the line element for the Reissner-Nordstr\"om-AdS
spacetime is
$ds^2 = - f\left(r\right) dt^2 +
\frac{dr^2}{f\left(r\right)} + r^2
\left( d\theta^2+\sin^2\theta\, d\phi^2\right)$,
with
$f\left(r\right) = \frac{r^2}{l^2} + 1 -
\frac{2M}{r} +\frac{Q^2}{r^2}$,
where
$M$ is the mass of the black hole
spacetime, $Q$ is its electric charge,
and 
$l^2 = - \frac3{\Lambda}$ is
the length scale related with the negative
cosmological constant $\Lambda$.
This line element can be rewritten in the form
\begin{equation}
ds^2 
\hskip-0.03cm
= 
\hskip-0.03cm
- 
\hskip-0.05cm
f\left(r\right) dt^2 
\hskip-0.03cm
+
\hskip-0.02cm
\frac{dr^2}{f\left(r\right)} 
\hskip-0.02cm
+
r^2 \left( d\theta^2
\hskip-0.05cm
+
\hskip-0.05cm
\sin^2\theta\,
d\phi^2\right),
\;\;\;
f(r)= \frac{1}{r^2}(r-r_+)(r-r_-)
\hskip-0.05cm
\left(\frac{r^2}{l^2}
\hskip-0.05cm
+ 
\hskip-0.05cm
\frac{r_+ 
\hskip-0.05cm
+ 
\hskip-0.05cm
r_-}{l^2}
r 
\hskip-0.05cm
+ 
\hskip-0.05cm
1
\hskip-0.05cm
+
\hskip-0.05cm
\frac{r_+^2 
\hskip-0.05cm
+ 
\hskip-0.05cm
r_-^2 
\hskip-0.05cm
+ 
\hskip-0.05cm
r_+ r_-}{l^2}
\hskip-0.05cm
\right),
\label{eq:fofr+r-s}
\end{equation}
where
the coordinate ranges are $-\infty<t<\infty$, $r_+<r<\infty$,
$0\leq\theta\leq\pi$, and $0\leq\phi<2\pi$, 
the radius $r_+$ is
the black hole event horizon radius, and
the radius $r_-$ is
the black hole Cauchy horizon radius.
The event horizon radius $r_+$
and the Cauchy horizon radius $r_-$
are given in terms of $M$, $Q$, and $l$ by
\begin{equation}
r_+ = M \sqrt{\mu} + \sqrt{M^2 \left[\left(\frac{1}{2\sqrt{\mu}} -
\frac{1}{3}\right) \frac{l^2}{M^2} - \frac{l^4}{12 M^4}
\frac{1}{\lambda^2} - \frac{\lambda^2}{12}\right] - Q^2
\frac{l^2}{M^2} \frac{1}{\lambda^2} }\,,
\label{eq:zeros_rplus}
\end{equation}
\begin{equation}
r_- = M \sqrt{\mu} - \sqrt{M^2 \left[\left(\frac{1}{2\sqrt{\mu}} -
\frac{1}{3}\right) \frac{l^2}{M^2} - \frac{l^4}{12 M^4}
\frac{1}{\lambda^2} - \frac{\lambda^2}{12}\right] - Q^2
\frac{l^2}{M^2} \frac{1}{\lambda^2} }\,,
\label{eq:zeros_rminus}
\end{equation}
where $\mu$ and $\lambda$ are defined as
\begin{equation}
\mu= \frac{l^4}{12 M^4} \frac{1}{\lambda^2} +\frac{\lambda^2}{12} +
\left(\frac{Q^2}{M^2} \frac{1}{\lambda^2} - \frac{1}{6}\right)
\frac{l^2}{M^2}\,,
\label{eq:mu_definition1}
\end{equation}
\begin{equation}
\lambda = \frac{1}{M}\biggl[l^6 + \left(54 M^2 -36 Q^2\right) l^4 +
\sqrt{108\left(M^2-Q^2\right)l^{10} + 108\left(27 M^4 -36 M^2 Q^2 + 8
Q^4\right)l^8 - 1728 Q^6 l^6}\biggr]^{\frac16}\,.
\label{eq:L_definition1}
\end{equation}

To obtain the pure Reissner-Nordstr\"om known expression
for $f(r)$ one has to perform
the zero cosmological constant limit
$\Lambda\to0$, i.e.,
$l \to \infty$ in Eq.~\eqref{eq:fofr+r-s}. This gives directly
that
$ \frac{r^2}{l^2} + \frac{r_+ + r_-}{l^2}
r + 1+\frac{r_+^2 + r_-^2 + r_+ r_-}{l^2}=1$ and so 
$f(r)= \frac{1}{r^2}(r-r_+)(r-r_-)$ as it should.
To obtain the pure Reissner-Nordstr\"om known expressions
for $r_+$ and $r_-$ in this limit
is slightly more involved.
All terms in Eqs.~\eqref{eq:zeros_rplus} and
\eqref{eq:zeros_rminus} have to be expanded to the same order.
To do that note that Eq.~\eqref{eq:mu_definition1} in this limit is
$\sqrt{\mu} = 1 + \mathcal{O}\left(\frac{1}{l^4}\right)$
and also that 
$\frac{l^2}{\sqrt{\mu}} = l^2 + 2
\left(2M^2-Q^2\right)+ \mathcal{O}\left(\frac{1}{l^2}\right)$.
In addition Eq.~\eqref{eq:L_definition1}
in this limit is
$\lambda^2 = \frac{l^2}{M^2} + 2 \sqrt{3}
\sqrt{M^2-Q^2}\frac{l}{M} + 6 +\mathcal{O}\left(\frac{1}{l}\right)$,
and also
$\frac{l^4}{\lambda^2} = M^2 l^2 - 2\sqrt{3} M^2 \sqrt{M^2-Q^2} l
+ 6M^2\left(M^2-2Q^2\right) +\mathcal{O}\left(\frac{1}{l}\right)$,
so that 
$\frac{l^2}{\lambda^2} = M^2 +\mathcal{O}\left(\frac{1}{l}\right)$.
Substituting these expansions in
the expressions appearing in Eqs.~\eqref{eq:zeros_rplus} and
\eqref{eq:zeros_rminus}, it can be shown that
$\left(\frac{1}{2\sqrt{\mu}} - \frac{1}{3}\right) \frac{l^2}{M^2} -
\frac{l^4}{12 M^4} \frac{1}{\lambda^2} - \frac{\lambda^2}{12} \to 1$,
$\frac{l^2}{\lambda^2 M^2} \to 1$, and $M\sqrt{\mu} \to M$.  Thus, the
horizon radii for a pure Reissner-Nordstr\"om black hole are
recovered in this limit,
i.e., $r_+ = M + \sqrt{M^2 - Q^2}$ and $r_- = M - \sqrt{M^2 -
Q^2}$.

\section{A compact
exact formula}
\label{interesting}

From
the equations of motion,
Eqs.~\eqref{eq:motion_t_penrose} and \eqref{eq:motion_r_penrose},
and 
the conservation laws,
Eqs.~\eqref{eq:energy_penrose}-\eqref{eq:X_penrose}, the
following identities can be derived for particle $i$,
where $i=1,2$,
\begin{equation}
 X_i = \frac{1}{2m_0^2} \left(X_0 b_i +
 P_0 \delta \sqrt{d}\right)\,,
    \label{eq:X12_penrose1app}
\end{equation}
\begin{equation}
 P_i = \frac{P_0 b_i - \left(-1\right)^i
 \delta X_0 \sqrt{d}}{2m_0^2}\,,
    \label{eq:X12_penrose2app}
\end{equation}
with  $b_i$, $d$, and $\delta$ being defined as
$b_i =
 m_0^2 - \left(-1\right)^i \left(m_1^2 -
 m_2^2\right)$,
$d =
 b_1^2 - 4m_0^2 m_1^2$,
and $\delta=\pm 1$.
The identities
given in Eqs.~\eqref{eq:X12_penrose1app}
and \eqref{eq:X12_penrose2app}
are the electrically charged analogous of the ones
derived for a rotating neutral metric in \cite{zaslavskii3}.

\section{Dynamics of odd particles in the decaying
scenarios 1 and 2}
\label{sec:odd_dynamics}

\subsection{Conditions for the motion of odd particles}

Odd particles move from the decaying point $r_{\rm i}$ in the inward
direction. To guarantee that odd particles fall into the
black hole, one has to show that motion is allowed all the way from
$r_{\rm i}$ to the black hole event horizon $r_+$, and that there are
no turning points along the way. To show this, we consider the
condition  which establish the radii where
particle motion is allowed, namely,
radii which satisfy $X^2 - m^2 f\left(r\right) \geq
0$, since the square root in Eq.~\eqref{eq:motion_r_penrose} has to
be non-negative. It is possible to rewrrite this condition
in terms of the
energy as
\begin{equation}
E\geq U(r)  \,,
    \label{eq:condition_epsilonapp}
\end{equation}
where the potential energy $U(r)$ is defined as
\begin{equation}
U(r)=m\sqrt{f\left(r\right)}+\frac{e Q}{r}   \,.
    \label{eq:condition_epsilon2app}
\end{equation}
In Eqs.~\eqref{eq:condition_epsilon} and
\eqref{eq:condition_epsilon2} we needed to use these
two equations for analyzing scenario 2.
Now we need to use them,
Eqs.~\eqref{eq:condition_epsilonapp} and
\eqref{eq:condition_epsilon2app}, 
for scenarios 1 and 2
applied to odd particles.
To prove
that there are no turning points between the horizon and the decaying
point, one has to examine if the equality in
Eq.~\eqref{eq:condition_epsilon} is never verified.

\subsection{Motion of odd particles in scenario 1}

Using the results for the electric charge and energy of odd particles,
Eqs.~\eqref{eq:charge1_case1_penrose} and
\eqref{eq:energyn1_case1_penrose2}, the condition in
Eq.~\eqref{eq:condition_epsilon} can be written as
$-\sqrt{f\left(r_{\rm i}\right)} \,\chi \geq \sqrt{f\left(r\right)} -
\left(1+\chi\right) \gamma \frac{Q}{r}$.
Rearranging this condition and using the definition
of $\gamma$ in Eq.~\eqref{eq:conditiondecay2}, one gets
\begin{equation}
\frac{r_{\rm i}}{r} + \chi \frac{r_{\rm i}-r}{r} \geq
\sqrt{\frac{f\left(r\right)}{f\left(r_{\rm i}\right)}}.
\label{eq:odd2}
\end{equation}
In the region $r_+ < r < r_{\rm i}$, the left hand side of
Eq.~\eqref{eq:odd2} is greater than 1, since $\frac{r_{\rm i}}{r} > 1$
and $\frac{r_{\rm i}-r}{r} >0$. The right hand side of the equation is
lower than 1, since $f\left(r\right)$ is a growing function of $r$ for
$r > r_+$ and we are considering a region where $r < r_{\rm i}$. Thus,
the inequality in Eq.~\eqref{eq:odd2} is always verified between the
decaying point and the event horizon. In fact, the equality is only
verified at $r = r_{\rm i}$. This means that $r_{\rm i}$ works as an
outer turning point for odd particles, which are always able to reach
the horizon in this decaying scenario. Therefore, odd particle indeed
fall into the black hole in scenario 1.

\subsection{Motion of odd particles in scenario 2}

Using now the expressions for the electric charge and energy of odd
particle in
the decaying scenario 2, Eqs.~\eqref{eq:charge2_case2_penrose}
and \eqref{eq:energy1_case2_penrose}, Eq.~\eqref{eq:condition_epsilon}
can be written as
$\sqrt{f\left(r_{\rm i}\right)} - \frac{e_0 Q}{r_{\rm i} m_0}
\frac{\lvert\beta_1\rvert}{\alpha_1}
\left(\frac{1-\beta_1}{1-\alpha_1}\right)^n \geq
\sqrt{f\left(r\right)} -\frac{e_0 Q}{r m_0}
\frac{\lvert\beta_1\rvert}{\alpha_1}
\left(\frac{1-\beta_1}{1-\alpha_1}\right)^n$,
which can be simplified into
\begin{equation}
\sqrt{f\left(r_{\rm i}\right)} - \sqrt{f\left(r\right)}
\geq -\frac{e_0 Q}{m_0}
\frac{\lvert\beta_1\rvert}{\alpha_1}
\left(\frac{1-\beta_1}{1-\alpha_1}\right)^n
\left(\frac{1}{r} - \frac{1}{r_{\rm i}}\right).
\label{eq:odd4}
\end{equation}
One finds that
$f\left(r_{\rm i}\right) > f\left(r\right)$. Thus, the left hand side
of Eq.~\eqref{eq:odd4} is positive. In the region between the event
horizon and the decay point, $\frac{1}{r} - \frac{1}{r_{\rm i}} > 0$,
which means that the right hand side of Eq.~\eqref{eq:odd4} is
negative. Therefore, the inequality in Eq.~\eqref{eq:odd4} is always
verified, which means that motion is allowed all the way into the
event horizon. One can also verify that the equality case is valid only
at $r = r_{\rm i}$, and so $r_{\rm i}$ is again an outer
turning point for odd particles, which fall into the black hole also
for this decaying scenario 2.

\section{Position of the outer turning point in the decaying
scenarios 1 and 2}
\label{sec:turning_position}

\subsection{General expression for the outer turning point}

Define
\begin{equation}
\tilde{E} = \frac{E}{m}\,,\quad\quad\quad\quad
\tilde{e} = \frac{e}{m}\,.
    \label{eq:unitmassdefs}
\end{equation}
Define
\begin{equation}
\bar{r} \equiv \frac{r}{M}\,,\quad\quad\quad\quad
\bar{r}_{\rm t} \equiv \frac{r_{\rm t}}{M}\,,\quad\quad\quad\quad
\bar Q \equiv \frac{Q}{M}\,,\quad\quad\quad\quad
\bar{l} = \frac{l}{M}\,.
    \label{eq:unitdefs}
\end{equation}

The position of the outer turning point $\bar{r}_{\rm t}$ obtained
as the largest real root of Eq.~\eqref{eq:turning_points_penrose} is
given by
\begin{equation}
\bar{r}_{\rm t} = \frac{1}{2\sqrt3}
\left( A +  \sqrt{4
\left(\tilde{E}^2 - 1\right) \bar{l}^2 - \frac{D}{3B^{\frac13}} -
\frac{B^{\frac13}}{2^{\frac13}} + \frac{12\bar{l}^2 \left(1
- \tilde{E} \tilde{e} \bar{Q}\right)}{A}}\right), 
    \label{eq:turning_point_position}
\end{equation}
where $A$, $B$, $C$, and $D$ are defined as
\begin{equation}
A =
\sqrt{2 \left(\tilde{E}^2 - 1\right) \bar{l}^2 +
\frac{D}{B^{\frac13}} + \frac{B^{\frac13}}{2^{\frac13}}},
\label{eq:definitions00}
\end{equation}
\begin{equation}
B = -2 \left(\tilde{E}^2-1\right)^3 \bar{l}^6 + 108 \bar{l}^4
\left(1-\tilde{E} \tilde{e} \bar{Q}\right)^2 - 72
\left(\tilde{E}^2 -1\right) \bar{l}^4 \bar{Q}^2
\left(\tilde{e}^2-1\right) - C,
\label{eq:definitions1}
\end{equation}
\begin{equation}
C = \sqrt{\hskip-0.05cm -4
\left[\left(\tilde{E}^2\hskip-0.05cm-
\hskip-0.05cm1\right)^2 \bar{l}^4
\hskip-0.05cm-\hskip-0.05cm12 \bar{l}^2 \bar{Q}^2
\left(\tilde{e}^2\hskip-0.05cm-
\hskip-0.05cm1\right)\right]^3\hskip-0.05cm
+\hskip-0.05cm \left[2\left(\tilde{E}^2-1\right)^3 \bar{l}^6
\hskip-0.05cm-\hskip-0.05cm 108 \bar{l}^4
\left(1\hskip-0.05cm-\hskip-0.05cm\tilde{E} \tilde{e}
\bar{Q}\right)^2\hskip-0.05cm +\hskip-0.05cm
72\left(\tilde{E}^2\hskip-0.05cm-\hskip-0.05cm1\right) \bar{l}^4
\bar{Q}^2
\left(\tilde{e}^2\hskip-0.05cm-\hskip-0.05cm1\right)\right]^2},
\label{eq:definitions2}
\end{equation}
\begin{equation}
D = 2^{\frac13} \left[\left(\tilde{E}^2-1\right)^2 \bar{l}^4
- 12
\bar{l}^2 \bar{Q}^2 \left(\tilde{e}^2-1\right)\right].
\label{eq:definitions3}
\end{equation}

\subsection{Large $n$ expansion for scenario 1}

In the large $n$ limit, the electric charge of even particles
in scenario 1 tends to
$e_\infty\equiv e_0+\left(1+\chi\right)\gamma \,m_0$,
according to Eq.~\eqref{eq:charge_2n_case1_penrose},
so that 
\begin{equation}
\tilde{e}_\infty \equiv \tilde{e}_0+
\left(1+\chi\right)\gamma\,.
\label{eq:einfty}
\end{equation}
In this limit the
energy tends to
$E_\infty\equiv E_0 + \gamma \chi \frac{{Q}}{r_{\rm i}}m_0$,
according to Eq.~\eqref{eq:energy_2n_case1_penrose},
so that 
\begin{equation}
\tilde{E}_\infty\equiv \tilde{E}_0 +
\gamma \chi \frac{{\bar Q}}{\bar r_{\rm i}}\,.
\label{eq:Einfty}
\end{equation}
Then, the charge per unit mass $\tilde{e}$
and the 
the energy per
unit mass $\tilde{E}$ are related in the form
$\tilde{e} = \frac{\tilde{e}_\infty}{\tilde{E}_\infty}\tilde{E}$.
Since the mass of even particles goes to zero along the
decaying
chain, the large $n$ limit implies large values for
the electric charge per unit mass
and for the energy
per unit mass. Therefore, the
position of the turning point can be expanded for large values of
$\tilde{E}$  as
\begin{equation}
\bar{r}_{\rm t} =
\frac{\tilde{e}_\infty \bar{Q}}{\tilde{E}_\infty} + 
\frac{1}{2}
\sqrt{2\bar{l}^2
\left(\frac{\tilde{E}_\infty}{\tilde{e}_\infty \bar{Q}}-1\right)
-\frac{4 \tilde{e}_\infty^2
\bar{Q}^2}{\tilde{E}_\infty^2}
}
\,.
\label{eq:xt_expanded_scenario1}
\end{equation}
This expression was obtained by
considering the following expansions
for $A$, $B$, $C$, and $D$, which for the sake
of completeness are given here in inline equations
as they can be taken from the definitions above,
$A = \frac{2 \tilde{e}_\infty \bar{Q}}{\tilde{E}_\infty}$,
$B = \left(- 2^{\frac13} \;\bar{l}^2\; \tilde{E}^2 + 2^{\frac13}
\left[ \bar{l}^2 + \frac{6 \tilde{e}_\infty^2
\bar{Q}^2}{\tilde{E}_\infty^2} - 2\sqrt{3}
\sqrt{\frac{B_1 + B_2}{r_{\rm i}^4\tilde{E}_\infty^4}}\right]\right)^3$,
with
$B_1 = Q \bar{l}^2 \left[\chi^2 \gamma^2
\bar{Q}^2 {\bar{r_{\rm i}}}^2
\left(6 \tilde{e}_\infty \tilde{E}_0 - 6 \tilde{E}_0^2
\bar{Q} -\tilde{e}_\infty^2
\bar{Q} \right) -\right.$ $\left.\chi^4 \gamma^4 \bar{Q}^5 - 2
\chi^3 \gamma^3
\bar{Q}^3 {\bar{r_{\rm i}}} (2 \tilde{E}_0 \bar{Q} -
\tilde{e}_\infty)\right]$
and
$B_2 = Q \bar{l}^2 \left[2 \chi \tilde{E}_0 \gamma \bar{Q}
{\bar{r_{\rm i}}}^3 \left(3 \tilde{e}_\infty \tilde{E}_0 -
\tilde{e}_\infty^2 \bar{Q} -
2 \tilde{E}_0^2 \bar{Q}\right) \!+ \!{\bar{r_{\rm i}}}^4
\!\left(2 \tilde{e}_\infty
\tilde{E}_0^3
\!- \!\tilde{e}_\infty^4
\frac{\bar{Q}^3}{\bar{l}^2}\!-\!\tilde{E}_0^2 \bar{Q}\times
\right.\right.$    
$\left.\left.\left(\tilde{e}_\infty^2+\tilde{E}_0^2\right)\right)\right]$,
$C = 12 \sqrt{3}\,\, \bar{l}^4 \sqrt{\frac{C_1 + 2\chi \gamma \tilde{E}_0
\bar{l}^2 \bar{Q}^2 {\bar{r_{\rm i}}}^3
\left(3\tilde{e}_\infty\tilde{E}_0-2\tilde{E}_0^2 \bar{Q}\right) +\bar{Q}
{\bar{r_{\rm i}}}^4 \left(2\tilde{e}_\infty \tilde{E}_0^3
\bar{l}^2 -\tilde{E}_0^2
\bar{l}^2
\bar{Q}\left(\tilde{e}_\infty^2+\tilde{E}_0^2\right)-
\tilde{e}_\infty^4
\bar{Q}^3\right)}{r_{\rm i}^4\tilde{E}_\infty^4}}$\hskip-0.05cm,
with $C_1$ given by
$C_1
\hskip-0.05cm
=
\hskip-0.05cm
2 \tilde{E}^4\chi^3 \gamma^3 \bar{l}^2 \bar{Q}^4 {\bar{r_{\rm i}}}
\left(\tilde{e}_\infty
\hskip-0.05cm
-
\hskip-0.05cm
2 \tilde{E}_0 \bar{Q}\right) + 6 \chi^2 \gamma^2
\bar{l}^2 \bar{Q}^3 {\bar{r_{\rm i}}}^2 \left(\tilde{e}_\infty
\tilde{E}_0-\tilde{E}_0^2\bar{Q}\right) - \chi^4 \gamma^4 \bar{l}^2
\bar{Q}^6$,
and
$D
\hskip-0.05cm
=
\hskip-0.05cm
2^{\frac13} \left[\bar{l}^4 \tilde{E}^4 \hskip-0.05cm - \hskip-0.05cm
\left(\frac{12 \tilde{e}_\infty^2 \bar{l}^2 \bar{Q}^2}{\tilde{E}_\infty^2}
\hskip-0.05cm
+
\hskip-0.05cm
2 l^4\right)
\hskip-0.05cm
\tilde{E}^2 \right.$
$\Bigl.+ \left(12 \bar{l}^2 \bar{Q}^2  +\bar{l}^4\right)
\Bigr]$.

\subsection{Large $n$ expansion for scenario 2}

In the large $n$ limit for decaying scenario 2, both $\tilde{E}$ and
$\tilde{e}$ take large values.
Therefore, the expression for the position of the
turning point can be expanded for large values of energy and electric
charge as
\begin{equation}
\bar{r}_{\rm t} = {\bar{r_{\rm i}}} + \frac{1}{2} \sqrt{-2 \bar{l}^2 -
4{\bar{r_{\rm i}}}^2 + \frac{2\bar{l}^2}{\bar r_{\rm i}} + 2\bar{l}^2
\sqrt{f\left({\bar{r_{\rm i}}}\right)} \tilde{E}}\,.
    \label{eq:xt_expanded}
\end{equation}
This expression was obtained by
considering the following expansions
for $A$, $B$, $C$, and $D$,  which for the sake
of completeness are given here in inline equations
as they can be taken from the definitions above,
$A = 2 {\bar{r_{\rm i}}}$,
$B \hskip-0.12cm = \hskip-0.15cm
\left(\hskip-0.15cm - 2^{\frac13} \bar{l}^2 \tilde{E}^2
\hskip-0.10cm + \hskip-0.10cm
\frac{2^{\frac13} \;\bar{l}^2 \left(\bar{l}^6+6 \bar{l}^4
{\bar{r_{\rm i}}}^2-2 \sqrt{3} \sqrt{-\bar{l}^8 \left(\bar{l}^2
\bar{Q}^2+\bar{l}^2 {\bar{r_{\rm i}}}^2-2 l^2 {\bar{r_{\rm i}}}-8 Q^4+9
{\bar{r_{\rm i}}}^4\right)}\right)}
{\bar{l}^6}\hskip-0.15cm\right)^{\hskip-0.15cm 3}\hskip-0.05cm$,
$C \hskip-0.12cm = \hskip-0.15cm
12 \sqrt{3} \tilde{E}^4 \sqrt{\hskip-0.05cm-\bar{l}^8 \left(\bar{l}^2
\bar{Q}^2+\bar{l}^2 {\bar{r_{\rm i}}}^2\hskip-0.05cm-\hskip-0.05cm 2 l^2
{\bar{r_{\rm i}}}
\hskip-0.05cm-\hskip-0.05cm8 Q^4+9
{\bar{r_{\rm i}}}^4\right)}$, and 
$
D = 2^{\frac13} \left[\bar{l}^4 \tilde{E}^4 - \left(2\bar{l}^4
+12 \bar{l}^2 {\bar{r_{\rm i}}}^2\right) \tilde{E}^2 + 24 \bar{l}^2
{\bar{r_{\rm i}}}^2
\sqrt{f\left({\bar{r_{\rm i}}}\right)} \tilde{E}
+ \left(12 \bar{l}^2\bar{Q}^2 \right.\right.$   $\left.\left.
- 12 \bar{l}^2 {\bar{r_{\rm i}}}^2 f\left({\bar{r_{\rm i}}}\right) +
\bar{l}^4\right)\right]$.

\section{Volume of the region between the event horizon and some
definite radius in the Reissner-Nordstr\"om-AdS spacetime}
\label{sec:volume}

The spatial volume between the black hole event horizon, $r_+$ and a
certain radius $r_{\rm f}$, in Reissner-Nordstr\"om-AdS spacetime, can be
obtained from the expression $V\left(r_{\rm f}\right) = 4\pi
\int_{r_+}^{r_{\rm f}} \frac{r^2}{\sqrt{f\left(r\right)}} \,dr$.
This integral can be expressed in terms of the variable $\bar{r}$
defined by 
\begin{equation}
\bar{r} =
\frac{r}{M}
\label{eq:barragain}
\end{equation}
as follows,
\begin{equation}
V\left(\bar{r}_{\rm f}\right) = 4\pi M^3
\int_{\bar{r}_+}^{\bar{r}_{\rm f}}
\frac{\bar{r}^2}{\sqrt{f\left(\bar{r}\right)}} \,d\bar{r},
\label{eq:volume_integral_x}
\end{equation}
where
$\bar{r}_+ =
\frac{r_+}{M}$,
$\bar{r}_{\rm f} =
\frac{r_{\rm f}}{M}$,
and 
$f\left(\bar{r}\right)$ can
be taken from Eq.~\eqref{eq:fofr+r-s}. To put 
$f\left(\bar{r}\right)$ in convenient terms
for our purposes of integration note that
the function that appears on the right hand side of 
Eq.~\eqref{eq:fofr+r-s}
can written in terms of the scaled variables as
$\frac{1}{{\bar l}^2} \bar{r}^2 + \frac{1}{{\bar l}^2}\left(\bar{r}_+ +
\bar{r}_-\right) \bar{r} + 1 + \frac{1}{{\bar l}^2} \left(\bar{r}_+^2 +
\bar{r}_-^2 + \bar{r}_+ \bar{r}_-\right)$, with
$\bar{r}_- =
\frac{r_-}{M}$ and ${\bar l}=\frac{l}{M}$.
It can also be written as 
$\frac{1}{{\bar l}^2}(\bar{r}-\bar{r}_{\rm c})(\bar{r}-\bar{r}_{\rm c}^*)=
\frac{1}{{\bar l}^2}\bar{r}^2-
\frac{1}{{\bar l}^2}(\bar{r}_{\rm c}+\bar{r}_{\rm
c}^*) \bar{r} + \frac{1}{{\bar l}^2}\bar{r}_{\rm c}\bar{r}_{\rm c}^*$, for
some $\bar{r}_{\rm c}$
and its complex conjugate $\bar{r}_{\rm c}^*$.  So, one deduces
$\bar{r}_{\rm c}+\bar{r}_{\rm c}^*=-\left(\bar{r}_+ +
\bar{r}_-\right)$, and 
$\bar{r}_{\rm c}\bar{r}_{\rm c}^*=\frac{1}{{\bar l}^2}
+ \bar{r}_+^2
+\bar{r}_-^2 + \bar{r}_+\bar{r}_-$.
Then,
$\bar{r}_{\rm c}$ can be found to be 
$\bar{r}_{\rm c} = - \frac{\bar{r}_+ + \bar{r}_-}{2} - i
\sqrt{\frac{3}{4} \left(\bar{r}_+^2+\bar{r}_-^2\right) + \frac{1}{2}
\bar{r}_+ \bar{r}_- + {\bar l}^2}$ and its complex conjugate
$\bar{r}_{\rm c}^*$
is of course given by $\bar{r}_{\rm c}^* = - \frac{\bar{r}_+ +
\bar{r}_-}{2} + i \sqrt{\frac{3}{4}
\left(\bar{r}_+^2+\bar{r}_-^2\right) + \frac{1}{2} \bar{r}_+ \bar{r}_-
+ {\bar l}^2}$. In this way, $f\left(\bar{r}\right)$
of Eq.~\eqref{eq:fofr+r-s} is 
\begin{equation}
f(\bar{r})=\frac{1}{{\bar l}^2\bar{r}^2}\left(\bar{r}-
\bar{r}_+\right)\left(\bar{r}-\bar{r}_-\right)
(\bar{r}-\bar{r}_{\rm c})(\bar{r}-\bar{r}_{\rm c}^*)
\,.
\label{eq:fox}
\end{equation} 
Performing then the integration of Eq.~\eqref{eq:volume_integral_x}
with the help of Eq.~\eqref{eq:fox},
the volume $V\left(\bar{r}_{\rm f}\right)$ is given by
\begin{equation}
V\left(\bar{r}_{\rm f}\right) =
2 \pi M^3 {\bar l}^2
\left(
\bar{r}_{\rm f}
\sqrt{f\left(\bar{r}_{\rm f}\right)} -
\frac{1}{\bar{r}_{\rm f} \sqrt{f\left(\bar{r}_{\rm f}\right)}}
h\left(\bar{r}_{\rm f}\right)
\biggl( \left(1-\bar{r}_-\right) F\left[u\left(\bar{r}_{\rm f}\right),
v\right] -\left(\bar{r}_+ - \bar{r}_-\right)
\Pi\left[w, u\left(\bar{r}_{\rm f}\right),
v\right]
\biggr)
\right),
\label{eq:volume_general}
\end{equation}
where
\begin{equation}
h\left(\bar{r}_{\rm f}\right) = \frac{2\left(\bar{r}_{\rm
f}-\bar{r}_{\rm c}^* \right)^2}{\left(\bar{r}_{\rm c}^* -
\bar{r}_+\right)^2} \sqrt{\frac{\left(\bar{r}_{\rm f}-\bar{r}_{\rm
c}^*\right) \left(\bar{r}_{\rm f}-\bar{r}_{\rm c}\right)
\left(\bar{r}_{\rm f}-\bar{r}_+\right)}{\left(\bar{r}_{\rm
c}-\bar{r}_-\right) }}\,,\quad\quad
u\left(\bar{r}_{\rm f}\right) = \arcsin\left(
\sqrt{\frac{\left(\bar{r}_{\rm f}-\bar{r}_{\rm c}\right)
\left(\bar{r}_{\rm c}^* -\bar{r}_+\right)}{\left(\bar{r}_{\rm f}-
\bar{r}_{\rm c}^*\right)
\left(\bar{r}_{\rm c}-\bar{r}_+\right)}}\right), 
\label{eq:volume_h}
\end{equation}
\begin{equation}
v = \frac{\left(\bar{r}_{\rm c}^* -
\bar{r}_-\right) \left(\bar{r}_{\rm c} -
\bar{r}_+\right)}{\left(\bar{r}_{\rm c} - \bar{r}_-
\right) \left(\bar{r}_{\rm c}^* -
\bar{r}_+\right)}\,,
\quad\quad\quad
w = \frac{\bar{r}_{\rm c} - \bar{r}_+}{\bar{r}_{\rm c}^* - \bar{r}_+},
\label{eq:volume_u_v_w}
\end{equation}
and
\begin{equation}
F\left[u\left(\bar{r}_{\rm f}\right), v\right] =
\int_{0}^{u\left(\bar{r}_{\rm f}\right)}
\frac{\,d\theta}{\sqrt{1-v\sin^2{\theta}}}\,,
\quad\quad\quad
\Pi\left[w; u\left(\bar{r}_{\rm f}\right), v\right] =
\int_{0}^{\sin{u\left(\bar{r}_{\rm f}\right)}}
\frac{1}{1-w t^2} \frac{\,dt}{\sqrt{\left(1-v
t^2\right)\left(1- t^2\right)}}\,,
\label{eq:firstthirdkind}
\end{equation}
are the incomplete elliptic integrals of the first
and third kind, respectively.
If wanted, the values $\bar{r}_+$, $\bar{r}_-$,
and $\bar{r}_{\rm c}$ can be explicitly 
given in terms of  ${\bar Q}=\frac{Q}{M}$
and ${\bar l}=\frac{l}{M}$.
From 
Eqs.~\eqref{eq:zeros_rplus} and \eqref{eq:zeros_rminus},
one has
$
\bar{r}_+ = \sqrt{\mu} + \sqrt{\left[\left(\frac{1}{2\sqrt{\mu}} -
\frac{1}{3}\right) {\bar l}^2 - \frac{{\bar l}^4}{12\lambda^2}
- \frac{\lambda^2}{12}\right] - \frac{{\bar Q}^2{\bar l}^2}{\lambda^2}}$,
and 
$
\bar{r}_- = \sqrt{\mu} - \sqrt{\left[\left(\frac{1}{2\sqrt{\mu}} -
\frac{1}{3}\right) {\bar l}^2 - \frac{{\bar l}^4}{12\lambda^2}
- \frac{\lambda^2}{12}\right] - \frac{{\bar Q}^2{\bar l}^2}{\lambda^2}}$,
with
$
\mu= \frac{{\bar l}^4}{12\lambda^2} +\frac{\lambda^2}{12} +
\left(\frac{{\bar Q}^2}{\lambda^2}
- \frac{1}{6}\right){\bar l}^2$, and
$
\lambda = \biggl[{\bar l}^6 + \left(54 -36 {\bar Q}^2\right) {\bar l}^4 +
\sqrt{108\left(1-{\bar Q}^2\right){\bar l}^{10} +
108\left(27  -36  {\bar Q}^2 + 8
{\bar Q}^4\right)l^8 - 1728 {\bar Q}^6 {\bar l}^6}\biggr]^{\frac16}$,
see Eqs.~\eqref{eq:mu_definition1}
and \eqref{eq:L_definition1}.
From $\bar{r}_{\rm c} = - \frac{\bar{r}_+ + \bar{r}_-}{2} - i
\sqrt{\frac{3}{4} \left(\bar{r}_+^2+\bar{r}_-^2\right) + \frac{1}{2}
\bar{r}_+ \bar{r}_- + {\bar l}^2}$ found above,
one then obtains,
in terms of  ${\bar Q}=\frac{Q}{M}$
and ${\bar l}=\frac{l}{M}$, that
$\bar{r}_{\rm c} = - \sqrt{\mu} - i
\sqrt{ \left[\left(\frac{1}{2\sqrt{\mu}} +
\frac{1}{3}\right) {\bar l}^2 + \frac{{\bar l}^4}{12\lambda^2}
+ \frac{\lambda^2}{12}\right] +
\frac{{\bar Q}^2{\bar l}^2}{\lambda^2}
}$
and its complex conjugate is
$\bar{r}_{\rm c}^* 
- \sqrt{\mu} +i
\sqrt{ \left[\left(\frac{1}{2\sqrt{\mu}} +
\frac{1}{3}\right) {\bar l}^2 + \frac{{\bar l}^4}{12\lambda^2}
+ \frac{\lambda^2}{12}\right] +
\frac{{\bar Q}^2{\bar l}^2}{\lambda^2}
}$
with $\mu$ and $\lambda$ as above.

We now calculate the particular case of the
volume of the region in Reissner-Nordstr\"om-AdS spacetime with
radius $r_{\rm t}$ in the
decaying scenario 2 when $n \to \infty$.
The volume of the region
between the black hole event horizon, $\bar{r}_+$, and the
turning point, $\bar{r}_{\rm f}$, given by Eq.~\eqref{eq:xt_expanded}
in the large $n$ limit, can be expressed as
\begin{equation}
V\left(\bar{r}_{\rm
t}\right) = 2 \pi M^3 \,\bar{l} \,\bar{r}_{\rm t}^2\,,
\label{eq:volume_rt}
\end{equation}
which grows with
$\bar{r}_{\rm t}^2$,
where it was used the result in
Eq.~\eqref{eq:volume_general}.  In terms of the radial coordinate $r =
\bar{r} M$ and of $l =
\bar{l} M$, the volume can be written as
$V\left(r_{\rm t}\right) = 2 \pi l r_{\rm t}^2$.
Note that this result can be obtained directly from
Eq.~\eqref{eq:volume_integral_x}, in the large $r_{\rm t}$ limit, since
$r_{\rm t}$ grows with $n$ for the decaying scenario 2. In this limit,
$f\left(r\right) = \frac{r^2}{l^2}$. Using this approximation in
Eq.~\eqref{eq:volume_integral_x}, the result
$V\left(r_{\rm t}\right) = 2 \pi l r_{\rm t}^2$
follows directly.



\begin{thebibliography}{999}


\bibitem{penrose_book}
R.~Penrose, ``Gravitational collapse: The role of general
relativity'', Rivista Nuovo Cimento {\bf 1} numero speciale,
252 (1969); reprinted in
Gen. Relativ. Gravit. {\bf 34}, 1141 (2002).

\bibitem{penrosefloyd}
R.~Penrose and R. M. Floyd, ``Extraction of rotational energy from a
black hole'', Nature Phys. Sci. {\bf 229}, 177 (1970).


\bibitem{bardeen}
J. M. Bardeen, W. H. Press, and S. A. Teukolsky, ``Rotating black
holes: Locally nonrotating frames, energy extraction, and scalar
synchrotron radiation'', Astrophys. J. {\bf 178}, 347 (1972).


\bibitem{wald}
R. M. Wald, ``Energy limits on the Penrose process'',
Astrophys. J. {\bf 191}, 231 (1974).

\bibitem{psk}
T. Piran, J. Shaham, and J. Katz, ``High efficiency of the Penrose
mechanism for particle collisions'', Astrophys. J. {\bf 196}, L107
(1975).

\bibitem{bsw} 
M. Ba\~nados, J. Silk, and S. M. West, ``Kerr black holes as particle
accelerators to arbitrarily high energy'', Phys. Rev. Lett. {\bf 103},
111102 (2009).

\bibitem{grib} 
A. A. Grib and Yu. V. Pavlov, ``On the collisions between particles in
the vicinity of rotating black holes'', Journal of Experimental and
Theoretical Physics JETP Letters {\bf 92}, 125 (2010); arXiv:1004.0913
[gr-qc].

\bibitem{gao} 
S. Gao and C. Zhong, ``Non-extremal Kerr black holes as particle
accelerators'', Phys. Rev. D {\bf 84}, 044006 (2011);
arXiv:1106.2852 [gr-qc].

\bibitem{zalavskii2}
O. B. Zaslavskii, ``On energetics of particle collisions near black
holes: BSW effect versus Penrose process'', Phys. Rev. D {\bf 86},
084030 (2012); arXiv:1205.4410 [gr-qc].

\bibitem{harada}
T. Harada, H. Nemoto, and U. Miyamoto, ``Upper limits of particle
emission from high-energy collision and reaction near a maximally
rotating Kerr black hole'', Phys. Rev. D {\bf 86}, 024027 (2012);
arXiv:1205.7088 [gr-qc].

\bibitem{haradakimura}
T. Harada and M. Kimura, ``Black holes as particle accelerators:
A brief review'', Classical Quantum Gravity {\bf 31},
243001 (2014); arXiv:1409.7502 [gr-qc].

\bibitem{schnittman}
J. D. Schnittman, ``The collisional Penrose process'', Gen.
Relativ. Gravit. {\bf 50}, 77 (2018); arXiv:1910.02800 [astro-ph.HE].

\bibitem{zaslavskii3}
O. B. Zaslavskii, ``Center of mass energy of colliding electrically
neutral particles and super-Penrose process'', Phys. Rev. D
{\bf 100}, 024050 (2019); arXiv:1904.04874 [gr-qc].

\bibitem{new}
F. Hejda, J. P. S. Lemos, and O. B. Zaslavskii, ``Extraction of energy
from an extremal rotating electrovacuum black hole: Particle
collisions in the equatorial plane'', Phys. Rev. D {\bf 105}, 024014
(2022); arXiv:2109.04477 [gr-qc].


\bibitem{newnew}
F. Hejda, J. P. S. Lemos, and O. B. Zaslavskii, ``Notes on extraction
of energy from an extremal Kerr-Newman black hole via charged particle
collisions'', Acta Phys. Pol. B Proc. Suppl. {\bf 15}, 1-A5 (2022);
arXiv:2202.13113 [gr-qc].

\bibitem{zalavskii7} 
O. B. Zaslavskii, ``On general properties of the Penrose process with
neutral particles in the equatorial plane'', Phys. Rev. D {\bf 108},
084022 (2023); arXiv:2307.06469 [gr-qc].



\bibitem{denardo}
G.~Denardo and R.~Ruffini, ``On the energetics of
Reissner-Nordst\"om
geometries'', Phys. Lett. B {\bf 45}, 259 (1973).



\bibitem{wdd}
S. M. Wagh, S. Dhurandhar, and N. Dadhich, ``Revival of the Penrose
process for astrophysical applications'',
Astrophys. J. {\bf 290}, 12 (1985).



\bibitem{zalavskii1}
O. B. Zaslavskii, ``Acceleration of particles by nonrotating charged
black holes'', Journal of Experimental and Theoretical Physics JETP
Letters {\bf 92}, 571 (2010); arXiv:1007.4598 [gr-qc].


\bibitem{zalavskii5} O. B. Zaslavskii,
``Energy extraction from extremal charged black
holes due to the BSW effect'',
Phys. Rev. D {\bf 86}, 124039 (2012); arXiv:1207.5209 [gr-qc].

\bibitem{nemoto}
H. Nemoto, U. Miyamoto, T. Harada, and T. Kokubu, ``Escape of
superheavy and highly energetic particles produced by particle
collisions near maximally charged black holes'', Phys. Rev. D {\bf 87},
127502 (2013); arXiv:1212.6701 [gr-qc].


\bibitem{zalavskii6} 
O. B. Zaslavskii, 
``Super-Penrose process for nonextremal black holes'',
Journal of Experimental and Theoretical Physics 
JETP Letters {\bf 113}, 757 (2021); arXiv:2103.02322 [gr-qc].


\bibitem{zeldovich}
Ya. B. Zel'dovich, ``Generation of waves by a rotating body'', Journal
of Experimental and Theoretical Physics JETP Letters {\bf 14}, 180
(1971).


\bibitem{starob}
A. A. Starobinsky, ``Amplification of waves during reflection from a
rotating black hole'', Journal of Experimental and Theoretical Physics
JETP {\bf 37}, 28 (1973).


\bibitem{teukpress1}
S. A. Teukolsky and W. H. Press, ``Perturbations of a rotating black
hole. III. Interaction of the hole with gravitational and
electromagnetic radiation'', Astrophys. J. {\bf 193}, 443 (1974).

\bibitem{dr}
T. Damour and R. Ruffini,
``Quantum electrodynamical effects in Kerr-Newmann geometries'',
Phys. Rev. Lett. {\bf 35}, 463 (1975).


\bibitem{teukpress2}
W. H. Press and S. A. Teukolsky, ``Floating orbits, superradiant
scattering and the black hole bomb'', Nature {\bf 238}, 211 (1972).

\bibitem{vilenkin}
A. V. Vilenkin, ``Exponential amplification of waves in the
gravitational field of ultrarelativistic rotating body'', Phys.
Lett. B {\bf 78}, 301 (1978).


\bibitem{cdly}
V.~Cardoso, O.~J.~C.~Dias, J.~P.~S.~Lemos, and S.~Yoshida, ``The black
hole bomb and superradiant instabilities'', Phys. Rev. D {\bf 70},
044039 (2004); arXiv:hep-th/0404096.


\bibitem{lemosanalogue}
J.~P.~S.~Lemos, ``Rotating analogue black holes: Quasinormal modes and
tails, superresonance, and sonic bombs and plants in the draining
bathtub acoustic hole'', in {\it Analogue Spacetimes: The First Thirty
Years}, Editors L. C. B. Crispino et al (Editora Livraria da F\'isica, S\~ao
Paulo, 2013), p.~145; arXiv:1312.7176 [gr-qc].

\bibitem{dhr}
J. C. Degollado, C. A. R. Herdeiro,
and H. F. R\'unarsson, ``Rapid growth of superradiant instabilities for
charged black holes in a cavity'', Phys. Rev. D
{\bf 88}, 063003 (2013); arXiv:1305.5513 [gr-qc].

\bibitem{dolanpw}
S. R. Dolan, S. Ponglertsakul, and E. Winstanley, ``Stability of black
holes in Einstein-charged scalar field theory in a cavity'',
Phys. Rev. D {\bf 92}, 124047 (2015); arXiv:1507.02156 [gr-qc].


\bibitem{menzay}
L. di Menza1, J.-P. Nicolas, and M. Pellen, ``A new type of charged
black hole bomb'', Gen. Relativ. Gravit. {\bf 52}, 8
(2020); arXiv:1903.02941 [gr-qc].


\bibitem{penrosekok}
T.~Kokubu, S.-L.~Li, P.~Wu, and H.~Yu, ``Confined Penrose process with
charged particles'', Phys. Rev. D {\bf 104}, 104047 (2021);
arXiv:2108.13768 [gr-qc].


\bibitem{penrosezasl}
O.~Zaslavskii, ``Confined Penrose process and black hole bomb'',
Phys. Rev. D {\bf 106}, 024037 (2022); arXiv:2204.12405 [gr-qc].




\bibitem{ddr}
T. Damour, N. Deruelle, and R. Ruffini, ``On quantum resonances in
stationary geometries'', Lett. Nuovo Cimento {\bf 15}, 257 (1976).


\bibitem{furuhashinambu}
H. Furuhashi and Y. Nambu, ``Instability of massive scalar fields in
Kerr-Newman spacetime'', Prog. Theor. Phys. {\bf 112}, 983 (2004);
arXiv:gr-qc/0402037.

\bibitem{cl}
V.~Cardoso and J.~P.~S.~Lemos, ``New instability for rotating black
branes and strings'', Phys. Lett. B {\bf 621}, 219 (2005);
arXiv:hep-th/0412078.

\bibitem{hawkingr}
S. W. Hawking and H. S. Reall, ``Charged and rotating AdS black holes
and their CFT duals'', Phys. Rev. D {\bf 61}, 024014 (1999);
arXiv:hep-th/9908109.

\bibitem{gubser}
S. S. Gubser, ``Breaking an Abelian gauge symmetry near a black hole
horizon'', Phys. Rev. D {\bf 78}, 065034 (2008); arXiv:0801.2977 [hep-th].

\bibitem{hartnoll}
S. A. Hartnoll, C. P. Herzog, and G. T. Horowitz, ``Building a
holographic superconductor'', Phys. Rev. Lett. {\bf 101}, 031601 (2008);
arXiv:0803.3295 [hepth].

\bibitem{uchikatayoshida}
N. Uchikata and S. Yoshida, ``Quasinormal modes of a massless charged
scalar field on a small Reissner-Nordström-anti-de Sitter black
hole'', Phys. Rev. D {\bf 83}, 064020 (2011); arXiv:1109.6737 [gr-qc].

\bibitem{green}
S. R. Green, S. Hollands, A. Ishibashi, and R. M. Wald, ``Superradiant
instabilities of asymptotically anti–de Sitter black holes'',
Classical Quantum Gravity {\bf 33}, 125022 (2016); arXiv:1512.02644
[gr-qc].

\bibitem{sanchis}
N. Sanchis-Gual, J. C. Degollado, P. J. Montero, J. A. Font, and
C. Herdeiro, ``Explosion and final state of an unstable
Reissner-Nordstr\"om black hole'', Phys. Rev. Lett. {\bf  116}, 141101
(2016); arXiv:1512.05358 [gr-qc].

\bibitem{bosch}
P. Bosch, S. R. Green, and L. Lehner, ``Nonlinear evolution and final
fate of charged anti–de Sitter black hole superradiant instability'',
Phys. Rev. Lett. {\bf 116}, 141102 (2016); arXiv:1601.01384
[gr-qc].

\bibitem{chandrasekhar} S. Chandrasekhar, {\it The Mathematical Theory
of Black holes} (Oxford University Press, Oxford, 1983).

\bibitem{olivaresetal}
M. Olivares, J. Saavedra, C. Leiva, and J. R. Villanueva, ``Motion of
charged particles on the Reissner-Nordström (anti)-de Sitter black
hole spacetime'', Mod. Phys. Lett. A {\bf 26}, 2923 (2011);
arXiv:1101.0748 [gr-qc].



\end{thebibliography}
\end{document}